\documentclass[12pt,preprint]{aastex}

\usepackage{epsfig}
\usepackage{pgfplots}
\usepackage{graphicx}
\usepackage{lscape}
\usepackage{amsmath}
\usepackage{amsbsy} 
\usepackage{rotating}
\usepackage{booktabs}
\usepackage{array}
\usepackage{lscape}
\usepackage{hyperref}
\usepackage{float}
\usepackage{amsfonts,amsmath,graphicx,lscape,nicefrac,url,hyperref,longtable,threeparttable}

\shorttitle{A new method for estimating the bolometric properties of Ibc SNe}

\begin{document}

\title{A new method for estimating the bolometric properties of Ibc SNe}
\author{Zach Cano\altaffilmark{1}}
\altaffiltext{1}{Centre for Astrophysics and Cosmology, Science Institute, University of Iceland, Reykjavik, Iceland, 107.}

\begin{abstract}

The bolometric properties (nickel mass, ejecta mass and kinetic energies) of 61 Ibc supernovae (SNe), including 20 Gamma-Ray Burst and X-Ray Flash (GRB/XRF), 19 Ib, 13 Ic and 9 Ic-BL (broad-lined) SNe are presented.  All of the available $BVRI$ photometry in the literature have been collected and used in a new method that utilizes a template supernova (SN 1998bw) and an analytical model based on Arnett (1982) to accurately estimate the bolometric properties of each SN.  A statistical analysis of the bolometric properties is then performed, where it is found that GRB/XRF SNe are the most energetic, and eject more mass (including nickel content) than Ib, Ic and Ic-BL SNe.  The results are then compared to the existing progenitor models of Ibc SNe, where it is concluded that it is highly likely that at least two progenitor channels exist for producing a Ibc SN: most Ibc SNe arise via binary interactions, where the mass of the stellar progenitor is less than what is attributed to a Wolf Rayet star.  Conversely, 
the progenitors of Ic-BL and GRB/XRF are more massive than those of Ib and Ic SNe, though a key difference between them is progenitor metallicity, with  Ic-BL SNe arise from more metal rich progenitors.  As mass loss in massive stars is influenced by metal content, the progenitors of Ic-BL SNe lose more mass, and therefore more angular momentum, before exploding.  It is expected that the explosion mechanism in Ic-BL and GRB/XRF SNe is ``engine-driven'' (i.e. an accreting black hole, or a millisecond magnetar), but the increased mass loss of Ic-BL SNe means the central engine is less powerful than in GRB/XRF SNe.  Finally, it is found that the SNe that accompany GRBs and XRFs are statistically indistinguishable, and some mechanism other than metallicity is needed to explain the differences in the high-energy components in these events.

\end{abstract}


\section{Introduction}

A core-collapse supernova (SN) marks the brilliant finale to the short life of a massive star.  The role of SNe in the shaping of the universe cannot be understated, especially when one considers their enormous input of ionizing photons, energy and chemical species into the universe.  It is expected that all stars greater than about 8 $\rm M_{\odot}$ will undergo a core-collapse at the end of their life (see Smartt 2009 for a review).  Many factors determine whether a SN may occur at all, and when it does, the exact type of supernova that will be produced, including: initial mass, chemical composition, rotation, magnetic fields, and whether the exploding star exists alone or is in a binary system.

Observationally there are two types of SNe, which are divided by the presence (type II) or lack of hydrogen (type I) in their peak spectra (see Filippenko 1997 for a review).  Type II SNe are further classified by the shape of the light curve (LC): most type II SNe possess a plateau in their LCs (IIP), while others demonstrate a linear decay after peak brightness (IIL).  Type I core-collapse supernovae (ccSNe) are further grouped according to the lack of additional chemical species in their spectra: Ib SNe possess helium features which are absent in the spectra of Ic.

Amidst this zoo of SNe, Ic-BL SNe are a unique subtype of Ic SNe that have been made famous by their association with long-duration gamma-ray bursts (GRBs) and x-ray flashes (XRFs).  Ic-BL SNe possess broad lines (BL) in their spectra, which indicate ejecta moving at very large velocities (in some cases more than 0.1$c$).  Indeed, since the first association of a GRB/XRF with a Ic-BL SNe was made in 1998, where SN 1998bw was seen to occur temporally and spatially coincident with GRB 980425 (Galama et al. 1998), all of the SNe associated with GRBs and XRFs have been Ic-BL.  However, there are many Ic-BL SNe that are not associated with a GRB or XRF.  These GRB/XRF-less Ic-BL SNe are seen to have ejecta moving at comparable velocities (e.g. Ic-BL SN 2009bb, Soderberg et al. 2010; Pignata et al. 2011), and are almost as energetic as GRB/XRF SNe, but have no corresponding gamma-ray emission.

GRB/XRF SNe\footnote{Here-on we refer to the Ic-BL SNe that are associated with GRBs and XRFs as GRB/XRF SNe, while the Ic-BL SNe that do not possess a gamma-ray emission component as Ic-BL SNe.} are rare events.  Several surveys, such as the Lick Observatory Supernova Search (LOSS; Smith et al. 2011), reveal that type II SNe are the most abundant ($\sim 75 \%$), while Ibc SNe, including all subtypes (e.g. Ic-BL and GRB/XRF SNe), account for only one in every four core-collapse SNe.  Furthermore, Ic-BL SNe are likely only 5\% of the total ccSNe population, and GRB/XRF SNe likely $10^{2}-10^{3}$ times less abundant than Ibc SNe (e.g. Podsiadlowski et al. 2004).  Thus a major goal in GRB/XRF-SN research is elucidating the unique and rare differences in the physical conditions and properties of the progenitor stars of GRBs and XRFs relative to those of other Ibc SNe.

Several clues have already been revealed.  The host environments of GRBs and XRFs have a lower metal content than those of Ib, Ic and Ic-BL SNe (Modjaz et al. 2008; Modjaz et al. 2011; Graham \& Fruchter 2012), with Ic SNe arising from more metal-rich environments than Ib and Ic-BL SNe (Modjaz et al. 2011). It is known that massive stars with higher metalicities will lose more mass via line-driven winds than massive stars with lower metalicities (e.g. Puls et al. 1996; Kudritzki et al. 2000; Mokeim et al. 2007), thus providing a natural explanation to why the progenitor stars of Ic SNe are more stripped of their outer envelopes than Ib SNe.

Indeed mass loss seems to play a major role in deciding whether a star will ultimately produce a GRB.  According to the ``collapsar'' model (see section \ref{sec:grb_vs_xrf}), an energetic jet cannot pierce through a star that possess an extended envelope (i.e. a supergiant), therefore the stellar progenitor needs to have a large amount of material removed for the jet to escape into space.  Such progenitors are expected to be Wolf-Rayet (WR) stars: highly evolved, stars whose outer envelopes of hydrogen and helium have been stripped away.  

But herein exists a paradox -- for while mass loss in massive stars comes via stellar winds, and the rate of which is highly dependent on metallicity, it is observed that GRBs and XRFs arise from metal-poor stellar progenitors.  So while the progenitors of GRBs and XRFs would undergo less mass loss before exploding, and therefore providing a way to retain the necessary angular momentum to then power a GRB, a mechanism is still needed to strip away the outer layers of the star.

Two solutions have been proposed.  The first are the quasi-chemically homogeneous stellar models of \cite{Yoon2005} and \cite{WoosleyHeger06}, which involve rapidly rotating single stars.  A consequence of the rapid rotation is that the stellar interior undergoes extensive mixing, meaning that the outer hydrogen envelope is almost entirely burned into helium.  This also has the effect that the stars do not undergo a giant phase in their evolution.  While some mass loss occurs in these models via stellar winds, it does so from a much more compact surface, thus allowing the stars to preserve enough angular momentum to then produce a GRB.

Whats more, quasi-chemically homogeneous stars appear to exist in nature.  The FLAMES survey (Hunter et al. 2008) observed over 100 O- and B-type stars in the Large Magellanic Cloud (LMC) and the Milky Way galaxy, and showed the presence of a group of rapidly rotating stars that were enriched with nitrogen at their surfaces.  The presence of nitrogen at the surface could only be due to rotationally triggered internal transport processes that brought nuclear processed material, in this case nitrogen, from the core to the stellar surface.  Other observations of metal-poor O-type stars in the LMC by \cite{Bouret2003} and \cite{Walborn2004} also show the signature of CNO cycle-processed material at their surfaces.

The second solution is if the stellar progenitors of GRBs and XRFs exist in binary systems.  Binary interaction offers a natural way for a star to be stripped of its outer envelope, most likely via Roche-lobe overflow, while the merger of the stars, possibly through a common-envelope phase (e.g. Chevalier 2012) can provide the means to generate the necessary angular momentum to power a GRB (Fryer \& Woosley 1998; Zhang \& Fryer 2001; Fryer \& Heger 2005).  There is a growing list of observations that show that most massive stars exist in binaries, including  \cite{Sana2012} who have estimated that over 70\% of all massive stars will exchange mass with a companion star, which in a third of all cases will lead to a merger of the binary system.

Additional support for the notion that the progenitors of Ibc SNe are massive stars in binary systems has come from \cite{Smith2011}, who have argued that for a standard initial mass function, the observed abundances of the different types of ccSNe are not consistent with expectations of single-star evolution.  Moreover, \cite{Eldridge2013} derived a 15\% probability that all Ibc SN arise from single-star Wolf-Rayet progenitors (i.e. a star that is highly stripped of its outer layers of hydrogen and helium).

To date, a major breakthrough in understanding the type of stars of ccSNe has arisen via direct imaging of the progenitor stars of type II SNe (see Smartt 2009 for a review).  However no such discoveries have been made regarding the progenitor stars of Ibc SNe (e.g. Eldridge et al. 2013).  Instead other means of ascertaining the properties of the progenitor stars has arisen mostly via modelling of spectroscopy and photometry of the SNe themselves (e.g. Iwamoto et al. 1998; Mazzali et al. 2002; Valenti et al. 2008).  While detailed modelling has revealed important clues to the properties of the stars that can give rise to certain Ibc SNe, the types of modelling vary from one event to another (spectroscopic vs. photometric).  Therefore comparing the results of these modelling is not always consistent from case to case.

Here I attempt to address this fact by performing a systematic analysis of published LCs of 61 Ibc SNe, including 20 GRB/XRF SNe and 9 Ic-BL SNe.  All of the available $BVRI$ photometry for these SNe have been gathered from the literature, and their stretch and luminosity factors in each filter relative to a k-corrected template supernova (SN 1998bw) was determined.  The average stretch and luminosity factors in all the available filters for each SN have been calculated, which were then used to transform the bolometric light curve (LC) of SN 1998bw.  The resultant bolometric LCs were then fit using an analytical model based on \cite{Arnett1982} to determine the nickel mass, explosion energy and ejecta mass of each SN.

The paper is arranged as follows.  In section \ref{sec:arnett} the Arnett model is discussed in detail, while in section \ref{sec:method} a detailed description of the method is given.  In section \ref{sec:grbcase_study} the validity of the method is demonstrated using the photospheric velocities obtained in section \ref{sec:photovel}, while in section \ref{sec:stats_analysis} the in-depth statistical analysis of the derived bolometric properties is presented.  A discussion of the implications of the results in given in section \ref{sec:discussion}, the caveats of the method are discussed in section \ref{sec:caveats}, and the conclusions are given in section \ref{sec:conclusion}.  The tables of photometry, photospheric velocities, stretch \& luminosity factors and bolometric properties of the entire sample of Ibc SNe can be found in the appendix.

Throughout the paper observer-frame times are used unless specified otherwise in the text.  Foreground reddening has been corrected for using the dust maps of \cite{Schlegel98}.  A flat $\Lambda$CDM cosmology with $H_{0}=71$ km s$^{-1}$ Mpc$^{-1}$, $\rm{\Omega_{M}} = 0.27$, and $\rm{\Omega_{\Lambda}=1-\Omega_{M}} =0.73$ is used.  

\section{The Arnett Model}
\label{sec:arnett}

\cite{Arnett1982} derived an analytical model for type I SNe - that is, SNe whose LCs are powered purely by the decay of radioactive elements.  This model has been used extensively since its initial formulation, as well as being built upon in recent years (e.g. \cite{Valenti08} extended the original model to include not only the energy produced by the decay of nickel into cobalt, but also the energy produced by the decay of cobalt into iron).  Interested readers are encouraged to study these papers to appreciate the more subtle intricacies of its theoretical derivation, and here the main facets of the model are given.  

At the heart of the the Arnett model are the following assumptions: 

\begin{enumerate}
\item A homologous expansion of the ejecta,
\item Spherical symmetry,
\item The radioactive nickel present in the ejecta is located at the centre of the explosion and does not mix,
\item Radiation-pressure dominance,
\item A small initial radius before explosion ($\rm{R_{o}} \rightarrow 0$),
\item The applicability of the diffusion approximation for photons (i.e. the \emph{photospheric phase}).  
\end{enumerate}

The luminosity of a type I SNe as a function of time is:

\begin{footnotesize}
\begin{center}
\begin{equation}
\label{equ:bol1}
\rm L(t) = M_{Ni}e^{-x^2} \times {\left((\epsilon_{Ni} - \epsilon_{Co}) \int_{0}^{x}A(z)dz+ \epsilon_{Co}\int_{0}^{x}B(z)dz\right)}
\end{equation} 
\end{center}
\end{footnotesize}

%\noindent
where 
\begin{center}
\begin{equation}
\label{equ:bol2}
A(z)=2ze^{-2zy+z^2}, B(z)=2ze^{-2zy+2zs+z^2} 
\end{equation} 
\end{center}
%\noindent
and $x\equiv t/\tau_{m}$, $y\equiv \tau_{m}/(2\tau_{Ni})$, and $s\equiv (\tau_{m}(\tau_{Co}-\tau_{Ni})/(2\tau_{Co}\tau_{Ni}))$.

The energy release in one second by one gram of $^{56}$Ni and $^{56}$Co are, respectively, $\epsilon_{\rm Ni}=3.90 \times 10^{10}$ erg s$^{-1}$ g$^{-1}$ and $\epsilon_{\rm Co}=6.78 \times 10^{9}$ erg s$^{-1}$ g$^{-1}$ (Sutherland \& Wheeler 1984; Cappellaro et al. 1997).  The decay times of $^{56}$Ni and $^{56}$Co, respectively, are $\tau_{\rm Ni}=8.77$ days (Taubenberger et al. 2006 and references therein) and  $\tau_{\rm Co}=111.3$ days (Martin 1987).  

$\tau_{m}$ is the effective diffusion time and determines the overall width of the bolometric light curve.  $\tau_{m}$ is expressed in relation to the  opacity $\kappa$ and the ejecta mass $\rm{M_{ej}}$, as well as the photospheric velocity ${v_{\rm ph}}$ at the time of bolometric maximum:

\begin{center}
\begin{equation}
\label{equ:tau} 
\tau_{m} \approx \left(\frac{\kappa}{\beta c}\right)^{1/2} \left(\frac{{M_{\rm ej}}}{{v_{\rm ph}}}\right)^{1/2}
\end{equation} 
\end{center}
%
%\noindent 
where $\beta \approx 13.8$ is a constant of integration (Arnett 1982), and $c$ is the speed of light. Additionally, we assume a constant opacity $\kappa=0.07$ cm$^{2}$g$^{-1}$ (e.g. Chugai 2000), which is justified if electron scattering is the dominant opacity source (e.g. Chevalier 1992).  And finally, the kinetic energy of the ejecta is simply ${E_{\rm k}} = \frac{1}{2} M_{\rm ej} v_{\rm ph}^{2}$.

\section{Method}
\label{sec:method}

The aim of this work is to obtain the bolometric properties of all of the SNe in our sample.  However, it is not possible to do this directly because of the lack of multi-wavelength observations for most of the events.  To compensate for this lack of data I have used a template supernova, SN 1998bw, for which there is a wealth of multi-wavelength observations.  This method therefore depends on the assumption that SN 1998bw is a good proxy for the SNe in our sample.  A detailed discussion of the caveats of this assumption is presented in section \ref{sec:caveats}.

The basic premise is then simple: to first determine the stretch ($s$) and luminosity ($k$) factors of the individual SNe relative to the k-corrected LCs of SN 1998bw in every available $BVRI$ filter, and then computed the average $s$ and $k$ for each SN. A simple empirical equation is then fit to the bolometric LC of SN 1998bw (equation \ref{equ:stretch}), and then the empirical relation is transformed by the average stretch and luminosity factors of each event using equation \ref{equ:stretch2}.  The transformed LC is then fit with the Arnett model, and using a value for the photospheric velocity (see section \ref{sec:photovel}), the bolometric properties of every SN in the sample are obtained.

Overall, our general method can be broken into two stages, first:

\begin{enumerate}
 \item Obtain LCs for each event; transform magnitudes into fluxes (using flux zeropoints in Fukugita et al. 1995); correct for foreground and rest-frame extinction; remove host contribution where necessary.
 \item Create a synthetic template light curve of SN 1998bw in each filter of how it would appear at the redshift of the SN being analysed.
 \item Fit the template LC with an empirical relation (equation \ref{equ:stretch}).
 \item Determine the relative stretch and luminosity factors in each filter to the fitted template (using equation \ref{equ:stretch2}).
 \item Calculate the average stretch and luminosity factors for each SN.
 \end{enumerate}

Once the stretch and luminosity factors have been determined in the individual filters, the following procedure was implemented to determine the bolometric properties of each SN:

\begin{enumerate}
\item Fit an empirical equation (equation \ref{equ:stretch}) to the $UBVRIJH$ bolometric LC of 1998bw (this needed to be done only once).
\item Apply the average stretch and luminosity factors to the empirical equation of 1998bw (using equation \ref{equ:stretch2}).
\item Using the Arnett model, determine the nickel mass, ejecta mass and kinetic energy of the SN. 
 \end{enumerate}

The last step can only be completed if the photospheric velocity of the SN ejecta is known.  For many events this is unknown, and so an \textit{average} velocity is used instead (see section \ref{sec:photovel}).

Our empirical relation is (Cano et al. 2011b):

\begin{center}
\begin{equation}
\label{equ:stretch}
U(t) = A + \lambda t\left(\frac{e^{(\frac{-t^{\alpha_{1}}} F)}}{1 + e^{(\frac{p-t} R) }}\right) + t^{\alpha_{2}}\log(t^{-\alpha_{3}})
\end{equation}
\end{center}

\noindent where $A$ is the intercept of the line and $\lambda$ and $F$ are related to the respective amplitude and width of the function.  The exponential cut-off function for the rise has a characteristic time $R$ and a phase zero-point $p$, while $\alpha_{1}$, $\alpha_{2}$ and $\alpha_{3}$ are free parameters.  All of the parameters are allowed to vary during the fit.  (\textit{NB: equation \ref{equ:stretch} is used throughout this investigation to fit the individual filter LCs as well as the bolometric LCs}.)

Next, the following relation was used to transform equation \ref{equ:stretch} by the average stretch and luminosity factors in Table \ref{table:grb_sk}:

\begin{equation}
\label{equ:stretch2}
 W(t)=k \times U(t/s)
\end{equation} 

Thus for events where it is not possible to construct a complete bolometric LC from individual filters, the average stretch and luminosity factors were used to transform equation \ref{equ:stretch} using equation \ref{equ:stretch2}.  Then equation \ref{equ:stretch2} is plotted along with the Arnett model to determine the bolometric properties of each of the Ibc SNe.

Ideally the foreground and rest-frame extinction needs to be determined, as well as the photospheric velocity for each event.  Proper knowledge of the total extinction along the line of sight is necessary to properly calculate the amount of nickel created during the explosion (as the overall peak brightness of the bolometric LC is determined by how much nickel is present in the ejecta).  The photospheric velocity is needed to determine the amount of material in the ejecta as well as the kinetic energy of the explosion.  For events where the rest-frame extinction is not known, the nickel masses derived in Tables \ref{table:grb_bol} and \ref{table:sn_bol} can be regarded as lower limits to the total nickel present.

The validity of the method depends on creating accurate k-corrected LCs of SN 1998bw at the redshift of each SN in the sample.  To do this a personally written program in C has been used.  The program calculates the rest-frame filters LCs for each SN and interpolates in frequency and time over the original LCs of SN 1998bw, and then uses the luminosity distance to rescale the flux density.  

GNUPlot was used to perform the fits, which uses a linear or non-linear least-squares fitting algorithm to minimize $\chi^{2}/ \rm dof$.  In the case of equation \ref{equ:stretch} and \ref{equ:stretch2}, the non-linear least-squares fitting algorithm is used.  The GNUPlot routine \textbf{fit} uses the Marquardt-Levenberg algorithm, which selects parameter values using an iterative process, and continues until the fit converges, i.e. when $\chi^{2}/ \rm dof$ is less than the pre-defined variable \textit{FIT\_LIMIT}.  The errors quoted by \textbf{fit} are parameter error estimates, i.e. ``standard errors'' rather than confidence intervals, and are calculated in the same way as the standard errors of a linear least-squares problem.  It is worth considering that the quoted errors by \textbf{fit} tend to be generally over-optimistic, but are useful for qualitative purposes.

The errors in the bolometric properties have been computed by considering the maximum and minimum stretch and luminosity factors, as well as the distribution of velocities around the average value (for those events where we have used an average ejecta velocity).

It must be stressed that we are only interested in determining the stretch and luminosity factors of the various SNe up to the first $25-30$ days after the explosion, i.e. during the ``photospheric phase'' for which the Arnett model is valid.  Therefore, to determine their peak and width relative to the template SN, the SN LCs were fit during this time frame only, and no attempt was made to ascertain a fit to the exponential tail of each SN relative to the template SN.

\section{Photospheric Velocities}
\label{sec:photovel}

The photospheric velocities of 35 Ibc SNe were collected from the literature: 7 GRB/XRF , 11 Ib, 10 Ic and 7 Ic-BL SNe (Table \ref{table:vel}).  Note that  all of the available photospheric velocities available in the literature were gathered, and not just the events in the sample, so to give extra statistical weight to the calculated average velocities.

From Table \ref{table:vel} we can see that Ib and Ic SNe have similar ejecta velocities.  It is seen that the average photospheric velocities of the Ib SNe is $v_{\rm ph,ave} = 8027$ km s$^{-1}$, while for the Ic SNe is $v_{\rm ph,ave} = 8470$ km s$^{-1}$.  The standard deviations of these averages are roughly $\sigma \sim 1700$ km s$^{-1}$.  Thus, when determining the bolometric properties of those Ib and Ic SNe that do not have a known photospheric velocity, an average velocity of $v_{\rm ph,Ibc} = 8000 \pm 2000$ km s$^{-1}$ is used, where the $\pm 2000$ km s$^{-1}$ reflects the distribution of velocities around this average value.

For the Ic-BL sample, there is a much wider spread in the photospheric velocities, and the value $v_{\rm ph,Ic-BL} = 15,000 \pm 4000$ km s$^{-1}$ has been adopted, which again reflects the distribution of velocities of this SN subtype.  Finally, for the seven GRB/XRF SNe that have measured ejecta velocities, a value of $v_{\rm ph,GRB/XRF} = 20,000 \pm 2500$ km s$^{-1}$ is used.

\section[]{GRB SNe case-studies}
\label{sec:grbcase_study}

The first step is to determine the validity of the method (i.e. if one can use the average stretch and luminosity factor of the individual $BVRI$ LCs as a proxy of the overall width and brightness of the bolometric LC).  To do this I investigated two case studies: XRF 100316D / SN 2010bh and XRF 060218 / SN 2006aj.  

For each case study the following steps were taken:

\begin{enumerate}
\item Determine the stretch and luminosity factors of SN 2010bh and SN 2006aj in $BVRI$ (which have been corrected for foreground and rest-frame extinction) relative to SN 1998bw. 
\item Take the stretch and luminosity factors in each filter and apply these to the corresponding $BVRI$ LCs of SN 1998bw.
\item Construct a $BVRI$ bolometric LC of SN 1998bw using the transformed LCs.
\item Construct a $BVRI$ bolometric LC of each case-study SN.
\item Construct a $BVRI$ bolometric LC of SN 1998bw using the original, untransformed LCs.
\item Fit a mathematical equation to the untransformed $BVRI$ bolometric LC of SN 1998bw (using equation \ref{equ:stretch}).
\item Transform this mathematical equation by the average stretch and luminosity factors in Table \ref{table:grb_sk} (using equation \ref{equ:stretch2}).
\item Using the Arnett model, fit the $BVRI$ bolometric LCs of SN 2010bh, SN 2006aj and the transformed bolometric LC of SN 1998bw and compare.
\end{enumerate}

\subsection{XRF 100316D / SN 2010bh}

\subsubsection{$BVRI$ Bolometric Light Curves}

\begin{figure}
 \centering
 \includegraphics[bb=0 0 280 188, scale=0.85]{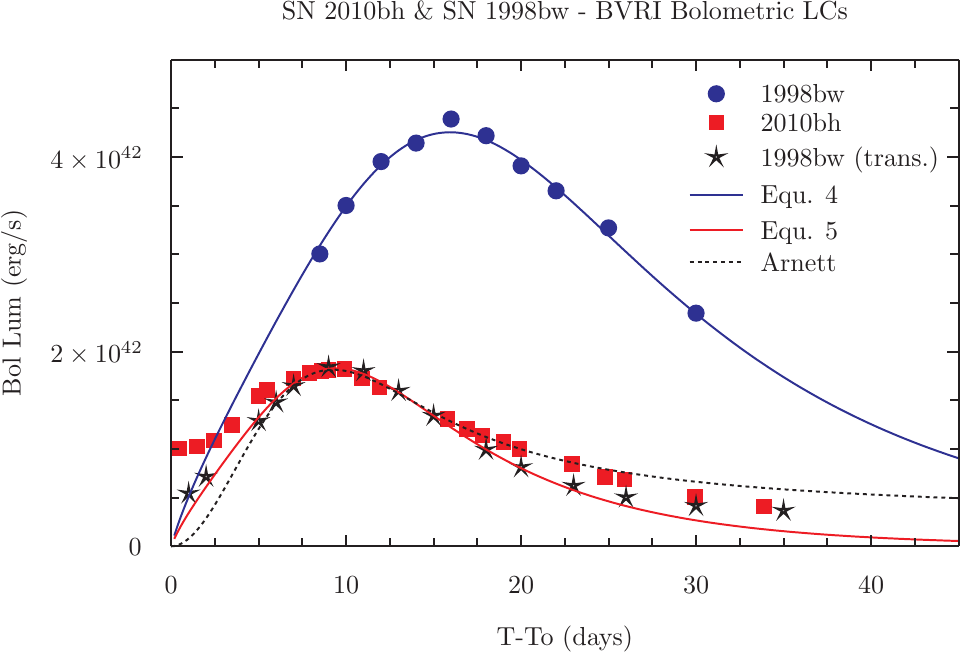}
 % 16D_BVRI.eps: 0x0 pixel, 300dpi, 0.00x0.00 cm, bb=-50 -31 231 158
\caption{$BVRI$ Bolometric Light Curves of SN 1998bw (blue circles) and SN 2010bh (red squares).  Also plotted is a transformed (black stars) $BVRI$ bolometric LC of SN 1998bw which has been transformed using the stretch and luminosity factors in $BVRI$ for SN 2010bh (Table \ref{table:grb_sk}).  Plotted alongside is equation \ref{equ:stretch} (blue line), which was fit to the untransformed bolometric LC of SN 1998bw, and equation \ref{equ:stretch2} (red line), which is equation \ref{equ:stretch} that has been transformed by the average stretch and luminosity factor for SN 2010bh ($s_{\rm ave,2010dh}=0.59$, $k_{\rm ave,2010dh}=0.43$). Finally, also plotted is the Arnett model (black, dotted line), which here has been fit to the transformed bolometric LC of SN 1998bw (see the text for the values obtained from fitting this model to the various bolometric LCs).}
\label{fig:16D_BVRI}
\end{figure}

Using the method outlined in the previous section, the hypothesis of whether a bolometric LC constructed from the transformed $BVRI$ LCs of SN 1998bw would appear the same as a bolometric LC constructed directly from the individual $BVRI$ LCs of SN 2010bh was tested.  Note that for SN 2010bh, the stretch and luminosity factors in each filter are determined by not including the early-time behaviour ($t-t_{0}<5$ days of the LCs).  \cite{Cano2011b} pointed out that the early LC contains an additional source of flux, which is likely due to light coming from shock-heated stellar material.  This behaviour was not seen in the optical LCs of SN 1998bw (Galama et al. 1998).  

All of the $BVRI$ bolometric LCs appear in Figure \ref{fig:16D_BVRI}.  Also plotted in this figure is equation \ref{equ:stretch}, which has been fit to the $BVRI$ bolometric LC of SN 1998bw, as well as equation \ref{equ:stretch2}, which has been transformed using the average stretch and luminosity factors for SN 2010bh ($s_{\rm ave,2010dh}=0.59$, $k_{\rm ave,2010dh}=0.43$; see Table \ref{table:grb_sk}).  Finally, the Arnett model is also plotted to fit the SN 2010bh bolometric LC.

Inspection of Figure \ref{fig:16D_BVRI} reveals that the transformed bolometric LC of SN 1998bw matches well to the original $BVRI$ bolometric LC of SN 2010bh after five days, as does equation \ref{equ:stretch2}.  When the Arnett model is fit to the $BVRI$ bolometric LC of SN 2010bh, and using a photospheric velocity of $v_{\rm ph}=25,000$ km s$^{-1}$ (Chornock et al. 2010) the following values are found: $M_{\rm Ni} = 0.051$ $\rm M_{\odot}$, $M_{\rm ej} = 2.34$ $\rm M_{\odot}$ and $E_{\rm k} = 1.45 \times 10^{52}$ erg.  When the model is applied to the transformed bolometric LC of SN 1998bw, it is found that $M_{\rm Ni} = 0.052$ $\rm M_{\odot}$, $M_{\rm ej} = 2.64$ $\rm M_{\odot}$ and $E_{\rm k} = 1.64 \times 10^{52}$ erg. 

Therefore, the relative differences between the original $BVRI$ bolometric LC of SN 2010bh and the transformed template are $M_{\rm Ni}$: 2\%, $M_{\rm ej}$: 13\% and $E_{\rm k}$: 13\%.

%\subsubsection{$UBVRI$ Bolometric Light Curves}

%Figure \ref{fig:16D_UBVRI} shows the $UBVRI$ bolometric LCs of 1998bw and 2010dh, as well as the Arnett model.  Also plotted is equation \ref{equ:stretch} that has been fit to the $UBVRI$ bolometric LC of 1998bw, as well as equation \ref{equ:stretch2} (using the average stretch and luminosity factors for 2010dh).  The results of the Arnett model for the bolometric LC of 2010dh are $M_{Ni} = 0.070 M_{\odot}$, $M_{ej} = 2.13 M_{\odot}$ and $E_{k} = 1.32 \times 10^{52}$ erg, while the results of the Arnett model applied to equation \ref{equ:stretch2} are $M_{Ni} = 0.078 M_{\odot}$, $M_{ej} = 2.20 M_{\odot}$ and $E_{k} = 1.37 \times 10^{52}$ erg.

\subsubsection{$UBVRIJH$ Bolometric Light Curves}

As we are most interested in determining the bolometric properties of our SNe over the widest filter range as possible, a $UBVRIJH$ bolometric LC of SN 2010bh as well as SN 1998bw has been constructed.  This time, as there are no accurate measurements of the stretch factors of SN 2010bh relative to SN 1998bw in $U$, $J$ and $H$, no attempt was made to construct a transformed template bolometric LC. 

\begin{figure}
 \centering
 \includegraphics[bb=0 0 280 188, scale=0.83]{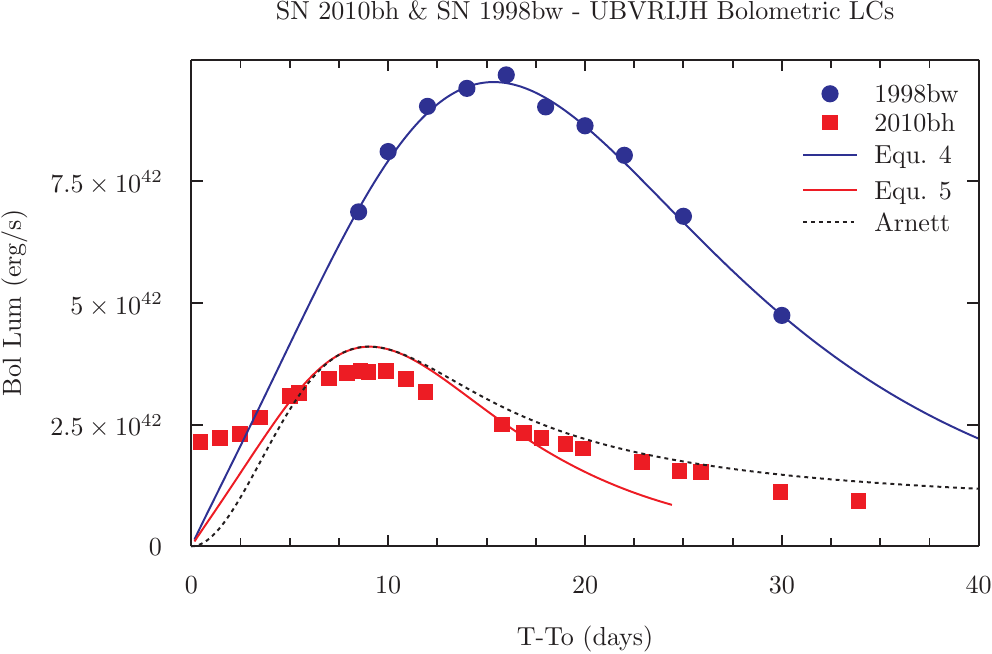}
 % 16D_BVRI.eps: 0x0 pixel, 300dpi, 0.00x0.00 cm, bb=-50 -31 231 158
\caption{$UBVRIJH$ Bolometric Light Curves of SN 1998bw (blue circles) and SN 2010bh (red squares).  Also plotted is equation \ref{equ:stretch} (blue line), which was fit to the untransformed bolometric LC of SN 1998bw, and equation \ref{equ:stretch2} (red line), which is equation \ref{equ:stretch} that has been transformed by the average stretch and luminosity factor for SN 2010bh ($s_{\rm ave,2010dh}=0.59$, $k_{\rm ave,2010dh}=0.43$).  Finally, also plotted is the Arnett model (black, dotted line), which here has been fit to equation \ref{equ:stretch2} (see the text for the values obtained from fitting this model to the various bolometric LCs).}
\label{fig:16D_UBVRIJH}
\end{figure}

Like before equation \ref{equ:stretch} was fit to the $UBVRIJH$ bolometric LC of SN 1998bw, and transformed this equation by the average stretch factors listed in Table \ref{table:grb_sk}.  Our results are displayed in Figure \ref{fig:16D_UBVRIJH}.  Also plotted is the Arnett model and the two empirical relations.

It is seen that equation \ref{equ:stretch2} has a slightly higher peak luminosity than the actual bolometric LC of SN 2010bh.  The results from fitting the Arnett model to the $UBVRIJH$ bolometric LC of SN 2010bh are $M_{\rm Ni} = 0.100$ $\rm M_{\odot}$, $M_{\rm ej} = 2.24$ $\rm M_{\odot}$ and $E_{\rm k} = 1.39 \times 10^{52}$ erg, while the results of the Arnett model applied to equation \ref{equ:stretch2} (i.e. the transformed bolometric LC of SN 1998bw) are $M_{\rm Ni} = 0.114$ $\rm M_{\odot}$, $M_{\rm ej} = 2.42$ $\rm M_{\odot}$ and $E_{\rm k} = 1.50 \times 10^{52}$ erg.  So while equation \ref{equ:stretch2} has a slightly higher peak luminosity, it amounts to only a 14\% difference in the estimates amount of nickel.  Only an 8\% difference was seen between the respective ejecta masses and kinetic energies.

%When we compare our results here with the published $UBVRIJH$ bolometric properties of SN 2010bh ($M_{\rm Ni} = 0.10 \pm 0.01$ $\rm M_{\odot}$, $M_{\rm ej} = 2.24 \pm 0.08$ $\rm M_{\odot}$ and $E_{\rm k} = 1.39 \pm 0.06 \times 10^{52}$ erg (Cano et al. 2011b)), we see there is excellent agreement with what we have derived here.

Thus, in both filter ranges, application of the Arnett model to the transformed bolometric LC of SN 1998bw reproduces very similar results to those obtained from directly fitting the bolometric LCs of SN 2010bh.  Moreover, despite the dissimilarities of the early LCs for each SN, we have been able to recover the bolometric properties of SN 2010bh.

\subsection{XRF 2006aj / SN 2006aj}

An identical analysis was also made using SN 2006aj.  This time a photospheric velocity of $v_{\rm ph}=20,000$ km s$^{-1}$ (Pian et al. 2006) is used.

\subsubsection{$UBVRI$ Bolometric Light Curves}

\begin{figure}
 \centering
 \includegraphics[bb=0 0 280 188, scale=0.85]{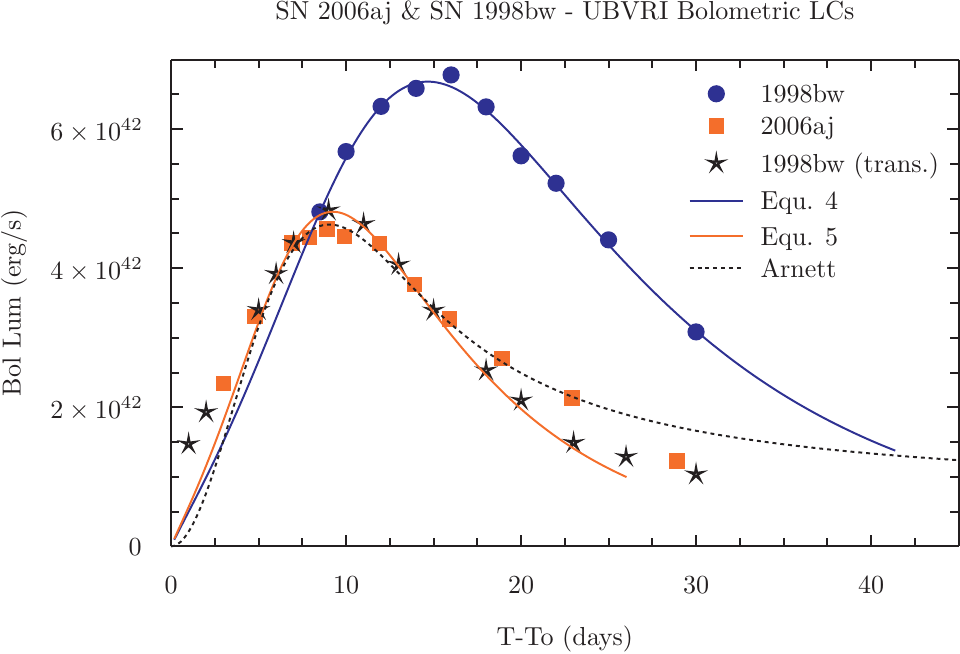}
 % 16D_BVRI.eps: 0x0 pixel, 300dpi, 0.00x0.00 cm, bb=-50 -31 231 158
\caption{$UBVRI$ Bolometric Light Curves of SN 1998bw (blue circles) and SN 2006aj (orange squares).  Also plotted is a transformed (black stars) $UBVRI$ bolometric LC of SN 1998bw which has been transformed using the stretch and luminosity factors in $UBVRI$ for SN 2006aj (Table \ref{table:grb_sk}). Plotted alongside is equation \ref{equ:stretch} (blue line), which was fit to the untransformed bolometric LC of SN 1998bw, and equation \ref{equ:stretch2} (red line), which is equation \ref{equ:stretch} that has been transformed by the average stretch and luminosity factor for SN 2006aj ($s_{\rm ave,2006aj}=0.65$, $k_{\rm ave,2006aj}=0.72$). Finally, also plotted is the Arnett model (black, dotted line), which here has been fit to the bolometric LC of SN 2006aj (see the text for the values obtained from fitting this model to the various bolometric LCs).}
\label{fig:2006aj_UBVRI}
\end{figure}

In this scenario, it was possible to accurately measure the stretch factor of SN 2006aj relative to SN 1998bw not only in $BVRI$ but also in $U$ (see Table \ref{table:grb_sk}).  As before, these stretch factors were applied to the individual $UBVRI$ LCs of the template SN, and a bolometric LC was constructed from the transformed LCs.  A $UBVRI$ bolometric LC of SN 2006aj was also constructed.  These bolometric LCs, as well as the untransformed bolometric LC of SN 1998bw, are displayed in Figure \ref{fig:2006aj_UBVRI}.  Also plotted is equation \ref{equ:stretch} that has been fit to SN 1998bw, as well as equation \ref{equ:stretch2}, using the average stretch and luminosity factor for SN 2006aj relative to SN 1998bw ($s_{\rm ave,2006aj}=0.65$, $k_{\rm ave,2006aj}=0.72$, see Table \ref{table:grb_sk}).  Finally, the Arnett model is also plotted to fit the bolometric LC of SN 2006aj.

The results of the Arnett model for the bolometric LC of SN 2006aj are $M_{\rm Ni} = 0.131$ $\rm M_{\odot}$, $M_{\rm ej} = 1.99$ $\rm M_{\odot}$ and $E_{\rm k} = 0.79 \times 10^{52}$ erg, while the results of the Arnett model applied to equation \ref{equ:stretch2} are $M_{\rm Ni} = 0.139$ $\rm M_{\odot}$, $M_{\rm ej} = 2.12$ $\rm M_{\odot}$ and $E_{\rm k} = 0.84 \times 10^{52}$ erg.  These correspond to relative differences of $M_{\rm Ni}$: 6\%, $M_{\rm ej}$: 6\% and $E_{\rm k}$: 6\%.

\subsubsection{$UBVRIJH$ Bolometric Light Curves}

Next, the $UBVRIJH$ bolometric properties of SN 2006aj were determined.  As before equation \ref{equ:stretch2} is plotted using the average stretch factors for SN 2006aj listed in Table \ref{table:grb_sk}, as well as the Arnett model.  It is found for the bolometric LC of SN 2006aj $M_{\rm Ni} = 0.178$ $\rm M_{\odot}$, $M_{\rm ej} = 2.05$ $\rm M_{\odot}$ and $E_{\rm k} = 0.82 \times 10^{52}$ erg, while for equation \ref{equ:stretch2} it is found $M_{\rm Ni} = 0.203$ $\rm M_{\odot}$, $M_{\rm ej} = 2.33$ $\rm M_{\odot}$ and $E_{\rm k} = 0.93 \times 10^{52}$ erg.  Again relative differences of $M_{\rm Ni}$: 13\%, $M_{\rm ej}$: 12\% and $E_{\rm k}$: 12\% are seen.

When these results are compared to the values found in the literature ($M_{\rm Ni} \approx 0.2$ $\rm M_{\odot}$, $M_{\rm ej} \approx 2$ $\rm M_{\odot}$ and $E_{\rm k} \approx 0.2 \times 10^{52}$ erg (Pian et al. 2006; Mazzali et al. 2006a)), good agreement is seen between the various permutations of the bolometric LC of SN 2006aj.  As noted in Cano et al. (2011b), when the Arnett model is used to analyse SN 2006aj, the kinetic energy is overestimated to that found by \cite{Pian06} and \cite{Mazzali2006a}.  An in-depth look at the caveats and limitations of the Arnett model are found in \cite{Cano2011b}, \cite{Cano12}, as well as in section \ref{sec:discussion}.

In summary, in both case studies and across three filter ranges, it is seen that it is possible to use the average stretch and luminosity factors as a proxy of the shape and brightness of the bolometric LC, and it is possible to recover the bolometric properties of both SN 2010bh and SN 2006aj using this method.  The average relative errors between the derived bolometric properties obtained from the transformed bolometric LCs to those obtained from the actual SN are of order 8-13\%  The next step is to repeat this procedure for the rest of the GRB SNe and then for the Ibc SNe.

\begin{table*}
\scriptsize
\centering
\setlength{\tabcolsep}{5pt}
\caption{Statistical Bolometric Properties of Ibc SNe}
\begin{tabular}{|c|cccc|cccc|cccc|cccc|}
\hline
&		&	$M_{\rm ej}$ 	&	$(\rm M_{\odot})$	&		&		&	$M_{\rm Ni}$$^{*}$ 	&	$(\rm M_{\odot})$	&		&		&	$E_{\rm k}$ 	&	$(10^{52}$ erg)	&		&		&	$E_{\rm k}/M_{\rm ej}$ 	&		&		\\
\hline
SN 	&	N	&	ave	&	stdev	&	median	&	N	&	ave	&	stdev	&	median	&	N	&	ave	&	stdev	&	median	&	N	&	ave	&	stdev	&	median	\\
\hline
Ib	&	19	&	4.72	&	2.77	&	3.89	&	12	&	0.21	&	0.22	&	0.16	&	19	&	0.33	&	0.26	&	0.23	&	19	&	0.066	&	0.023	&	0.064	\\
Ic	&	13	&	4.55	&	4.51	&	3.40	&	7	&	0.23	&	0.19	&	0.19	&	13	&	0.33	&	0.33	&	0.22	&	13	&	0.071	&	0.024	&	0.064	\\
Ibc	&	32	&	4.65	&	3.51	&	3.56	&	19	&	0.21	&	0.21	&	0.18	&	32	&	0.33	&	0.28	&	0.22	&	32	&	0.068	&	0.023	&	0.064	\\
Ic-BL	&	9	&	5.42	&	3.44	&	3.90	&	6	&	0.36	&	0.33	&	0.26	&	9	&	1.26	&	0.89	&	1.09	&	9	&	0.245	&	0.106	&	0.224	\\
GRB	&	14	&	6.34	&	4.40	&	5.39	&	14	&	0.40	&	0.27	&	0.34	&	14	&	2.49	&	1.77	&	2.14	&	14	&	0.390	&	0.029	&	0.398	\\
XRF	&	6	&	5.57	&	2.46	&	6.45	&	6	&	0.36	&	0.19	&	0.33	&	6	&	2.14	&	0.72	&	2.37	&	6	&	0.413	&	0.111	&	0.397	\\
GRB/XRF	&	20	&	6.11	&	3.87	&	5.91	&	20	&	0.39	&	0.24	&	0.34	&	20	&	2.38	&	1.52	&	2.23	&	20	&	0.397	&	0.063	&	0.398	\\
\hline
\end{tabular}
\label{table:Ibc_bolo_stats}
\begin{flushleft}
$^{*}$  Only those events where the rest-frame/host extinction is know have been included. \\
\end{flushleft}

\end{table*}

\subsection{The rest of the Ibc SNe}

It was possible to determine the stretch and luminosity factors of the GRB/XRF SNe using equation \ref{equ:stretch2} because the exact time of explosion was known (i.e. $t$ in equation \ref{equ:stretch2} is actually $t-t_{0}$, where $t_{0}$ is the explosion time, which for GRB/XRF SNe is the time the GRB pulse is detected).  However, for the non-GRB/XRF SNe in our sample, the time of explosion is unknown.  To account for this, equation \ref{equ:stretch2} has been altered so that $t \equiv t-t_{\rm peak}$ instead (i.e. the stretch and luminosity factors are now being found relative to the $peak$ of the LC).  Thus the following equation was fit to the Ibc SNe to determine $s$, $k$ and $t_{\rm peak}$:

\begin{equation}
\label{equ:stretch3}
 W(t)=k \times U((t-t_{\rm peak})/s)
\end{equation} 

Using the method described in the preceding section, along with equation \ref{equ:stretch3}, the stretch and luminosity factors of the complete sample of Ibc SNe relative to SN 1998bw have been obtained, and the results are displayed in the appendix (GRB/XRF SNe in Table \ref{table:grb_sk}; and in Table \ref{table:sn_ks} for the Ib, Ic and Ic-BL SNe).  Then, using the photospheric velocities of the Ibc SNe (Table \ref{table:vel}), either those determined from the individual spectra, or using an average value (see section \ref{sec:photovel}), the bolometric properties of each Ibc SN in the sample have bee calculated.  The results of the modelling are also displayed in the appendix (GRB/XRF SNe: Table \ref{table:grb_bol}; Ib, Ic, Ic-BL in Table \ref{table:sn_bol}).

\section{Statistical Analysis}
\label{sec:stats_analysis}

\subsection{General properties of Ibc SNe}

Using the ejecta masses, nickel masses and explosion energies in Tables \ref{table:grb_bol} and \ref{table:sn_bol}, the average and median values for each Ibc SN subtypes have been calculated (Table \ref{table:Ibc_bolo_stats}).  

Although the sample sizes of bolometric properties of Ibc SNe, including GRB/XRF SNe, presented here is the largest yet considered, there are a few events (e.g. GRB 041006; SN 2011bm) that affect the statistics significantly.  For example, the average ejecta masses for Ic SNe is 4.55 $\rm M_{\odot}$, with a standard deviation of the distribution being 4.51, and the median value being 3.40 $\rm M_{\odot}$.  However, when SN 2011bm is removed from the Ic sample, the average drops to 3.36 $\rm M_{\odot}$ (i.e. very close the median value of the sample when SN 2011bm is included in it), and has a standard deviation of 1.53.  Thus the removal of a single event drastically affects the average statistics.  To account for the effect that single events can have on the sample statistics, the median values for each of the bolometric properties have been calculated, and it is these that are consider in the following analysis.  

Several conclusions can be drawn from the analysis:

\begin{itemize}
 \item Median ejecta masses:
 \begin{itemize}
    \item Ib: $M_{\rm ej} \sim 3.9$ $\rm M_{\odot}$
    \item Ic: $M_{\rm ej} \sim 3.4$ $\rm M_{\odot}$
    \item Ic-BL: $M_{\rm ej} \sim 3.9$ $\rm M_{\odot}$.
    \item GRB/XRF: $M_{\rm ej} \sim 6.0$ $\rm M_{\odot}$.
\end{itemize} 
 \item Median nickel masses:
  \begin{itemize}
    \item Ib and Ic: $M_{\rm Ni} \sim 0.15-0.18$ $\rm M_{\odot}$.
    \item Ic-BL: $M_{\rm Ni} \sim 0.25$ ${\rm M_\odot}$.
    \item GRB/XRF: $M_{\rm Ni} \sim 0.3-0.35$ $\rm M_{\odot}$.
\end{itemize} 
\item Median kinetic energies
  \begin{itemize}
    \item Ib and Ic: $E_{\rm k} \sim 0.2 \times 10^{52}$ erg.
    \item Ic-BL: $E_{\rm k} \sim 1.0 \times 10^{52}$ erg.
    \item GRB/XRF: $E_{\rm k} \sim 2.0 \times 10^{52}$ erg.
\end{itemize} 
 \item Kinetic energy - ejecta mass ratios
  \begin{itemize}
    \item Ib and Ic: $E_{\rm k}/ M_{\rm ej} \sim 0.06$ ($10^{52}$ erg $\rm M_{\odot}^{-1}$).
    \item Ic-BL: $E_{\rm k}/ M_{\rm ej} \sim 0.2$ ($10^{52}$ erg $\rm M_{\odot}^{-1}$).
    \item GRB/XRF: $E_{\rm k}/ M_{\rm ej} \sim 0.4$ ($10^{52}$ erg $\rm M_{\odot}^{-1}$).
\end{itemize} 
 
\end{itemize}

\begin{table*}
%\tiny
\centering
\setlength{\tabcolsep}{5pt}
\caption{KS Test - results}
  \begin{tabular}{|c|cc|cc|cc|}
  \hline
	&	$M_{\rm ej}$	&		&	$M_{\rm Ni}$ 	&		&	$E_{\rm k}$ 	&		\\
	\hline
	\hline
SN pair	&	D	&	$p$	&	D	&	$p$	&	D	&	$p$	\\
\hline
Ib, Ic	&	0.1857	&	0.9098	&	0.2211	&	0.9417	&	0.2071	&	0.8262	\\
-	&	-	&	-	&	-	&	-	&	-	&	-	\\
Ib,Ic-BL	&	0.2000	&	0.9238	&	0.4065	&	0.3436	&	0.7500	&	0.0004	\\
Ic,Ic-BL	&	0.2412	&	0.9197	&	0.2321	&	0.9702	&	0.7571	&	0.0010	\\
-	&	-	&	-	&	-	&	-	&	-	&	-	\\
GRB,Ib	&	0.3666	&	0.1546	&	0.5230	&	0.0273	&	0.8166	&	$6.11 \times 10^{-06}$	\\
GRB,Ic	&	0.5238	&	0.0230	&	0.4833	&	0.1204	&	0.7904	&	0.0001	\\
GRB,Ic-BL	&	0.3666	&	0.3170	&	0.3238	&	0.6060	&	0.5666	&	0.0248	\\
-	&	-	&	-	&	-	&	-	&	-	&	-	\\
XRF,Ib	&	0.4214	&	0.2386	&	0.4945	&	0.1488	&	0.8071	&	0.0008	\\
XRF,Ic	&	0.5000	&	0.1320	&	0.3392	&	0.6851	&	0.7857	&	0.0024	\\
XRF,Ic-BL	&	0.3714	&	0.5145	&	0.2857	&	0.8827	&	0.6142	&	0.0512	\\
-	&	-	&	-	&	-	&	-	&	-	&	-	\\
GRB/XRF,Ib	&	0.3666	&	0.0969	&	0.5412	&	0.0103	&	0.8547	&	$1.44\times 10^{-07}$	\\
GRB/XRF,Ic	&	0.5238	&	0.0118	&	0.4642	&	0.1152	&	0.8095	&	$9.64\times 10^{-06}$	\\
GRB/XRF,Ic-BL	&	0.3666	&	0.2554	&	0.3333	&	0.5123	&	0.6142	&	0.0062	\\
-	&	-	&	-	&	-	&	-	&	-	&	-	\\
GRB/XRF,Ibc	&	0.4242	&	0.0132	&	0.4690	&	0.0141	&	0.8138	&	$1.92\times 10^{-08}$	\\
-	&	-	&	-	&	-	&	-	&	-	&	-	\\
GRB,XRF	&	0.2380	&	0.9116	&	0.1999	&	0.9795	&	0.1999	&	0.9795	\\

\hline
\end{tabular}
\label{table:KStest}
\begin{flushleft}
D is the maximum vertical deviation between the two empirical distributions, and $p$ is the probability that the two samples are drawn from the same parent population.  One is able to reject the null hypothesis (i.e. that there is no relationship between two measured phenomena) when $p$ is small (usually taken to be $p<0.05$).

\end{flushleft}
\end{table*}

\begin{figure}
 \centering
 \includegraphics[angle=270,bb=50 50 554 770,scale=0.7,clip=true,trim=150pt 0pt 0pt 0pt]{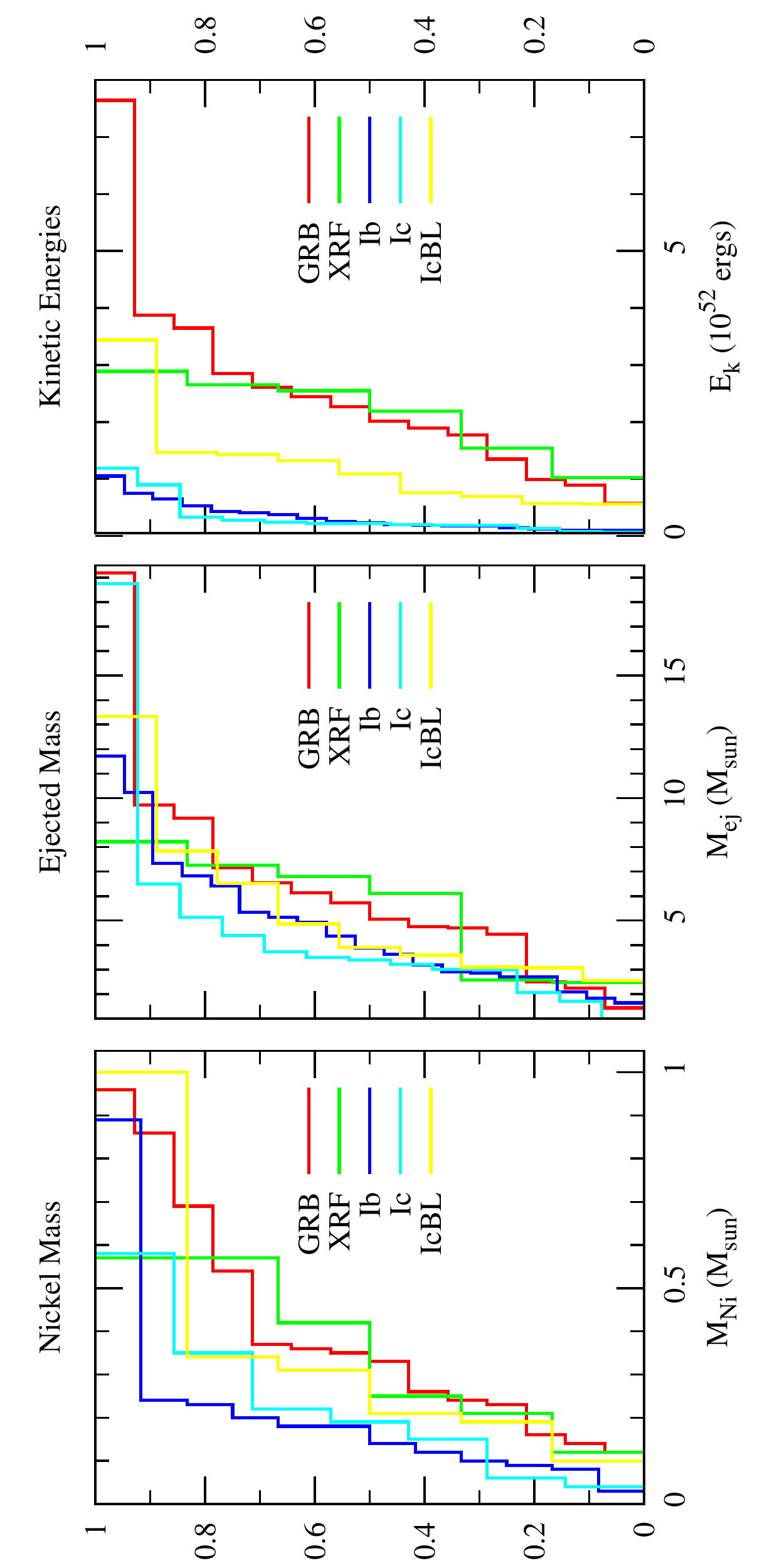}
 % test.eps: 0x0 pixel, 300dpi, 0.00x0.00 cm, bb=50 50 554 770
 \caption{Cumulative frequency plots of the nickel mass, ejecta masses and kinetic energies of the Ibc SNe.}
\label{fig:cumm_freq}
\end{figure}

One particularly striking result is a clear trichotomy of the kinetic energies seen for Ibc, Ic-BL and GRB/XRF SNe.  This can also be seen in Figure \ref{fig:cumm_freq}.  However, as the kinetic energies calculated here is simply ${E_{\rm k}} = \frac{1}{2} M_{\rm ej} v_{\rm ph}^{2}$, it could be argued that the different average values arise from the fact that an average velocity has been used for those events that didn't have a known ejecta velocity.  To check if this is the case, an investigation was made for only those events where the ejecta velocity has been determined from spectra.  This reduces the sample sizes to: Ibc (11), Ic-BL (6) and GRB/XRF (7).  The median values for the three subgroups are virtually identical to those seen for the complete sample ($E_{\rm k} = 0.18, 1.04, 2.01 \times 10^{52}$ erg for Ibc, Ic-BL and GRB/XRF respectively), and confirms that the trichotomy is real.

Similarly for the kinetic energy/ejecta mass ratios, a clear trichotomy is seen.  To check whether using an average ejecta velocity causes this trichotomy, I have re-considered only those events where the ejecta velocity has been determined via spectroscopy.  Again the median values for the subgroups are very similar to the entire sample: $E_{\rm k}/ M_{ej} \sim 0.06, 0.21, 0.39$ ($10^{52}$ erg $\rm M_{\odot}^{-1}$) for Ibc, Ic-BL and GRB/XRF respectively.

While the sample sizes of some of the Ibc subtypes are relatively small (i.e. we only have 9 Ic-BL SNe in the sample), though those of the Ibc and GRB/XRF SNe are quite large (32 and 20 respectively), what is encouraging is that the same trends are seen in both the average and median values for the bolometric properties of each subtype.  No matter which statistic is considered, the conclusion that GRB/XRF SNe eject more mass, including nickel content, than Ibc and Ic-BL SNe does not change.  Similarly for the explosion energies -- a clear trichotomy exists in the average and median values.  More encouraging is that fact that in the larger samples (i.e. the complete Ibc and GRB/XRF samples), the median and average values tend to converge towards a common value.  This hold more for the GRB/XRF SNe sample than the Ibc sample, where a clear difference still exists between the average and median ejecta masses and explosion energies, but is less pronounced in the nickel masses.

\subsection{Kolmogorov-Smirnov test}

Using the derived bolometric properties, a Kolmogorov-Smirnov (KS) test\footnote{in the KS-test, D is the maximum vertical deviation between the two empirical distributions, and $p$ is the probability that the two samples are drawn from the same parent population.  One is able to reject the null hypothesis (i.e. that there is no relationship between two measured phenomena) when $p$ is small (usually taken to be $p<0.05$).} was performed, using a personally written python script, to determine whether any of the Ibc SN subtypes are drawn from the same parent population (i.e. arise from the same type of progenitor stars) based on the distribution of each subtype's bolometric properties.  The results of the tests are found in Table \ref{table:KStest}.  A cumulative frequency plot of the ejecta masses, nickel masses and kinetic energies is presented in Figure \ref{fig:cumm_freq}.

The results of the KS-test show that when the largest samples (Ibc vs. GRB/XRF) are considered, one is able to reject the hypothesis that these two populations of SNe are drawn from the same parent population.  This is based on the $p$ values obtained between the two samples for each of the bolometric properties.

When the results of the KS-test are considered for the kinetic energies of Ibc, Ic-BL and GRB/XRF SNe, it is also possible to reject the hypothesis that all three are drawn from the same parent population, i.e. when considering the following SN pairs: GRB/XRF vs Ic-BL, Ibc vs. Ic-BL and GRB/XRF vs. Ibc.  As seen in the previous section, a clear trichotomy is seen in the average kinetic energies of each subtype, and the results of the KS-test also support this conclusion.

population.

When considering these results one must remain mindful that the KS test is very sensitive to sample size.  While the entire Ibc sample, excluding the Ic-BL SNe, is $N=32$, and the GRB/XRF SN sample is $N=20$, the individual Ibc samples (e.g. Ic, Ic-BL, XRF SNe) are still very small.  Therefore, when considering small sample(s), it is not surprising that the test doesn't reject any assumption, simply because the low sample sizes imply very low sensitivity of the test.

\subsection{Comparison with published values}

While an inspection of Table \ref{table:grb_bol} and \ref{table:sn_bol} shows that the bolometric properties that have been estimated here are consistent with those published in the literature, an element of caution is needed when comparing the values found here with those in the literature, especially those that have been estimated by the authors using an Arnett-like model.  If we consider equation \ref{equ:tau} again, it can be re-written as:

\begin{center}
\begin{equation}
\label{equ:tau2}
M_{\rm ej} \approx \left(\frac{\beta c}{\kappa}\right) \times v_{\rm ph} \times \tau_{\rm m}^{2} \times \lambda
\end{equation} 
\end{center}

\noindent Here I have introduced an additional variable, $\lambda$, which in this work is $\lambda \equiv 1$.  In other works $\lambda$ varies, depending on the version of equation \ref{equ:tau} that the authors use.  For example, \cite{Valenti08} have $\lambda = \sqrt{3/20} = 0.387$, while \cite{Taubenberger06} have $\lambda = 0.5$, and \cite{Pignata2011} have $\lambda = 1$.  Moreover, different values for the optical opacity are used by different authors.  For example, here $\kappa = 0.07$ cm$^{2}$ g$^{-1}$ is used, while \cite{Pignata2011} and \cite{Valenti08} use $\kappa = 0.08$, and \cite{Drout2011} use $\kappa = 0.05$.  Thus different values for $\lambda$ and $\kappa$ will affect the final estimates of the ejected mass and kinetic energies of the SNe (see Cano et al. 2011b and Cano 2012 for further discussions).  

It is worth noting that the different values of $\lambda$ and $\kappa$ will not affect the \textit{relative} average ejected masses and kinetic energies between the different subtypes as all the SNe will be affected the same way.\footnote{Assuming that the opacity in all of the Ibc SNe is the same, which for the Ic, Ic-BL and GRB/XRF SNe may be correct, however this may be less correct for Ib SNe where there is a moderate amount of helium in the ejecta.  See the caveats in section \ref{sec:caveats} for a further discussion.}  Therefore, for whichever values of $\lambda$ and $\kappa$ are used, one will always find that GRB/XRF SNe eject the most mass and are the most energetic, while Ic-BL are intermediate, and Ibc eject the least amount of material and are the least energetic.  Note also that the nickel masses are not affected by the values of $\lambda$ or $\kappa$, and are determined only by the peak luminosity of the SN.  

With this caveat in mind, when a comparison is made to the values found in the literature for the average ejecta masses of smaller samples of Ibc SNe, higher average and median values are found in this paper.  For example, \cite{Drout2011} find for their sample of Ibc SNe, average ejecta masses of: Ib = $2.0^{+1.1}_{-0.8}$ $\rm M_{\odot}$, Ic: $1.7^{+1.4}_{-0.8}$ $\rm M_{\odot}$, and Ic-BL: $4.7^{+2.3}_{-1.8}$ $\rm M_{\odot}$.  The values of ejecta masses for the Ibc SNe is much higher here (we find that the ejecta masses for Ib and Ic SNe are twice that found by \cite{Drout2011}), both for the average and median values, respectively, (Ib: 4.72, 3.89 $\rm M_{\odot}$; Ic: 4.55, 3.40 $\rm M_{\odot}$; Ibc: 4.65, 3.56 $\rm M_{\odot}$), though the values for the Ic-BL SNe do overlap (5.42, 3.90 $\rm M_{\odot}$).  However, the difference in the ejecta masses between Ibc and Ic-BL SNe that was seen by \cite{Drout2011} is not seen here, though if we take the average ejecta masses at face value they imply that Ic-BL 
eject marginally more mass than Ib and Ic SNe.  This difference is not seen in the median values however, which may be more appropriate for the sample sizes we are considering here. 

\cite{Cano12} collected from the literature a smaller sample of ejecta masses for Ibc SNe, including Ic-BL and GRB/XRF SNe (his Tables 5.10 and 5.11).  The tabulated ejecta masses were derived via different methods (photometric and spectroscopic modelling), and the average ejecta masses are similar to what is found here:  Ib $\sim 4.3$ $\rm M_{\odot}$, Ic $\sim 2.4$ $\rm M_{\odot}$, Ic-BL $\sim 5.1$ $\rm M_{\odot}$ and GRB/XRF $\sim 7.1$ $\rm M_{\odot}$.  In all cases, whether in this work or comparing literature values, the conclusions do not change.	

It is also worth noting that here average ejecta velocities of $8,000$ km s$^{-1}$ and $15,000$ km s$^{-1}$ are used for Ibc and Ic-BL SNe respectively, while \cite{Drout2011} uses $10,000$ km s$^{-1}$ and $20,000$ km s$^{-1}$.  Using different velocities will affect the estimates of the ejecta mass and kinetic energies, thus further muddying the comparison between the two samples.

\begin{table*}
%\scriptsize
\centering
\setlength{\tabcolsep}{5pt}
\caption{High-energy and metallicity properties of GRB/XRF SNe}
  \begin{tabular}{|ccccccc|}
  \hline
type	&	Name	&	$z$	&	$E_{\rm p}$ (keV)	&	$E_{\rm iso}$ ($10^{52}$ erg)	&	$\rm 12 + log(O/H)$ (dex)$^{*}$ 	&	Ref.	\\
\hline		
\hline
XRF	&	980425	&	0.0085	&	$122\pm17$	&	$0.00009$	&	8.16	&	(1),(2)	\\
XRF	&	020903	&	0.25	&	$3.37\pm1.79$	&	$0.0024\pm0.0006$	&	8.22	&	(2),(3)	\\
XRF	&	031203	&	0.1055	&	$>190$	&	$0.017$	&	8.02	&	(1),(2)	\\
XRF	&	060218	&	0.0331	&	$4.9\pm0.3$	&	$0.0053\pm0.0003$	&	7.54	&	(3),(4)	\\
XRF	&	100316D	&	0.0591	&	$18\pm3$	&	$0.006$	&	8.20	&	(1),(5)	\\
XRF	&	120422A	&	0.283	&	$53$	&	$0.0045$	&		&	(1)	\\
GRB	&	990712	&	0.434	&	$93\pm15$	&	$0.67\pm0.13$	&	8.10	&	(2),(3)	\\
GRB	&	991208	&	0.706	&	$313\pm31$	&	$22.3\pm1.8$	&	8.73 (7.40)	&	(2),(3)	\\
GRB	&	011121	&	0.36	&	$1060\pm265$	&	$7.8\pm2.1$	&	8.64 (7.50)	&	(2),(3)	\\
GRB	&	020405	&	0.69	&	$354\pm10$	&	$10\pm0.9$	&	8.44 (7.78)	&	(2),(3)	\\
GRB	&	021211	&	1.006	&	$46.8\pm5.8$	&	$0.66$	&		&	(1)	\\
GRB	&	030329	&	0.1685	&	$100\pm23$	&	$1.5\pm0.3$	&	7.97 (8.33)	&	(2),(3)	\\
GRB	&	041006	&	0.716	&	$98\pm20$	&	$3\pm0.9$	&		&	(3)	\\
GRB	&	050525A	&	0.606	&	$127\pm10$	&	$2.5\pm0.43$	&		&	(3)	\\
GRB	&	050824	&	0.83	&		&	1	&		&	(6)	\\
GRB	&	060729	&	0.5428	&		&	1.6	&		&	(7)	\\
GRB	&	060904B	&	0.703	&	$\sim 85$	&	$\sim 2$	&		&	(8)	\\
GRB	&	080319B	&	0.937	&	$1261\pm65$	&	$114\pm9$	&		&	(3)	\\
GRB	&	090618	&	0.54	&	$155.5\pm11$	&	$25.7$	&		&	(9),(10)	\\
GRB	&	091127	&	0.49	&	$45$	&	$1.1\pm0.2$	&		&	(11),(12)	\\
\hline													
\hline													
Redshift	&	Subgroup	&	average	&	stdev	&	D	&	$p$	&		\\
\hline
	&	XRF	&	0.123	&	0.116	&	0.8666	&	0.0005	&		\\
	&	GRB	&	0.624	&	0.225	&	-	&	-	&		\\
	&	all	&	0.473	&	0.306	&	-	&	-	&		\\
\hline													
\hline													
Metalicity	&	Subgroup	&	average	&	stdev	&	D	&	$p$	&		\\
\hline													
(upper)	&	XRF	&	8.03	&	0.28	&	0.5000	&	0.3180	&		\\
	&	GRB	&	8.38	&	0.33	&	-	&	-	&		\\
	&	all	&	8.20	&	0.34	&	-	&	-	&		\\
-	&	-	&	-	&	-	&	-	&	-	&		\\
(lower)	&	XRF	&	8.03	&	0.28	&	0.3333	&	0.8095	&		\\
	&	GRB	&	7.82	&	0.39	&	-	&	-	&		\\
	&	all	&	7.93	&	0.34	&	-	&	-	&		\\
\hline													
										
\end{tabular}
\label{table:GRB_HE}

\medskip
\begin{flushleft}
$^{*}$ Metalicity using the upper (lower) estimates from \cite{Savaglio09}.\\
(1) \cite{Zhang2012}, (2) \cite{Savaglio09}, (3) \cite{Amati2008}, (4) \cite{Wiersema2007}, (5) \cite{Levesque10}, (6) \cite{Sollerman2007}, (7) \cite{Grupe2007}, (8) \cite{Margutti2008}, (9) \cite{Ghirlanda2010}, (10) \cite{Page2011}, (11) \cite{Troja2012}, (12) \cite{Berger2011}
\end{flushleft}

\end{table*}

\section{GRB SNe vs XRF SNe}
\label{sec:grb_vs_xrf}

In the preceding sections it was seen that the average bolometric properties of GRB and XRF SNe are statistically indistinguishable.  This is in contrast to the high-energy properties of GRB and XRF SNe, where many differences clearly exist.  GRBs have an isotropic energy release in gamma rays of $E_{\rm iso} \sim 10^{51}-10^{54}$ erg, while XRFs emit much less gamma radiation: $E_{\rm iso} \sim 10^{47}-10^{49}$ erg (see Table \ref{table:GRB_HE}).  Moreover, the shape of the gamma-ray pulse in XRFs tend to be very smooth and non-variable, while those seen for GRBs can exhibit variability on the order of milliseconds (e.g. Kaneko et al. 2007; Gehrels et al. 2009).  Further differences exist in the x-ray LCs: in XRFs 060218 and 100316D the x-ray LCs show an initial plateau (lasting a few 1000 sec), that is then followed by a sudden and very steep decay (e.g. see Figure 7 in Starling et al. 2011).  This is quite different to the x-ray LCs seen in cosmological GRBs that tend to follow a ``steep-shallow-steep'' 
temporal evolution over the same timescales (e.g. Nousek et al. 2006; Zhang et al. 2006; Evans et al. 2009).  Such different observational behaviour hints at different physical processes generating the gamma-ray and x-ray radiation in GRBs and XRFs.

Theoretically, the generally accepted progenitor scenario for producing a GRB and GRB SN is the ``collapsar'' model (e.g. Woosley 1993, MacFayden \& Woosley 1999, MacFayden et al. 2001).  In this model a stellar black hole (BH) is formed when the collapse of a massive star fails to produce a strong supernova explosion. If material falling back and accreting upon the BH has sufficient angular momentum, it can hang up and form a disk.  Accretion from the disk onto the BH is then thought to produce a relativistic, bi-polar jet along its rotation axis which punctures through the stellar material and generates a burst of gamma rays at a large distance from the star, where the optical depth is low and the gamma rays can escape into space.  The explosion mechanism in GRB SNe is then due to the interaction of the jet with the stellar material as it pierces through the star.  An alternative model is the ``millisecond magnetar'' model (e.g. Usov 1992), where the energy source for powering a GRB and the accompanying SN 
also arises from the collimated outflow/jet of a compact object, but instead of a stellar BH, it arises from a strongly magnetized, rapidly rotating neutron star.

However, observations of many XRF SNe appear to be at odds with the collapsar and millisecond-magnetar scenarios.  Instead, work (e.g. MacFayden et al. 2001, Bromberg et al. 2011, Bromberg et al. 2012, Nakar \& Sari 2012) has suggested that XRFs are produced by ``failed jets'' - jets that are produced via an accreting BH, but are weak and fail to break out of their progenitor stars.  The failed jet dissipates all of its energy into the star and it is this that drives the SN explosion.  Therefore in XRF SNe, the weak gamma-ray emission arises from the relativistic shock-breakout and not via the same mechanism that produces a GRB. 

One reason why the jet may fail to break out of the stellar progenitor in XRFs is that the stars that produce XRFs have a higher metal content than those of GRBs.  There is observational evidence that shows that the explosion sites of GRBs are more metal poor than those of other types of Ibc and Ic-BL SNe (e.g. Modjaz et al. 2008; Modjaz et al. 2011; Graham \& Fruchter 2012), however no evidence exists as to whether the progenitor stars of XRFs have a higher metal content to those of GRBs.  As mass loss in massive stars is largely dependant on metallicity, if XRFs are more metal rich than GRBs then more angular momentum will lost by the BH and accretion disk.  Thus the BH would be rotating more slowly in XRFs, implying a weaker energy source.  \cite{WoosleyZhang2007} showed that below a certain energy, $\sim 10^{48}$ erg, a jet cannot bore through the star and escape into space.

To investigate whether a metallicity difference exists in the environments of GRBs and XRFs, all of the metallicity measurements made for the host galaxies of the GRB and XRF SNe in our sample were assembled from the literature to determine whether any statistically significant differences exist between them (Table \ref{table:GRB_HE}).  

The analysis displayed at the bottom of Table \ref{table:GRB_HE} clearly shows that there is no statistically significant difference between the average metalicities of the hosts and sites of GRBs and XRFs.  This is seen when both the upper and lower metallicity estimate from \cite{Savaglio09} are considered (their Table 9).  A KS test of the distributions also supports this conclusion -- that it is likely that both populations arise from progenitors of similar metallicities.  

\cite{GrahamFruchter2012} have also compiled a sample of GRB and XRF metallicities from the literature (their Table 1), which contain 4 XRFs and 10 GRBs.  Using their metallicity values, which have been computed from the tabular line fluxes via the Kobulnicky \& Kewley (2004) R23 diagnostic, one calculates average metallicities of: GRB $=8.45$ dex and XRF $=8.36$ dex.  This result supports the conclusion that GRBs and XRFs arise from environments with similar amounts of metal content.

Next, the redshift distributions of GRB and XRF SNe was determined, and it is seen that the average redshifts of the two samples are very different: for XRF and GRB SNe $z=0.123, 0.627$ respectively.  This is expected as it is impossible to detect XRFs in the distant universe due to their intrinsically low gamma-ray emission.  

Therefore, the result that GRBs and XRFs arise from environments of similar metallicity is perhaps unexpected if galaxy metallicity evolves with redshift  (i.e. galaxies at higher redshifts have lower metal content) -- one would then expect the environments of GRBs to be more metal poor as they arise from a population of stars with a higher average redshift.  However, the recent results of \cite{GrahamFruchter2012} have shown that the low metallicity distribution of GRBs is \textit{not} caused by the general properties of their host galaxies.  Moreover, they excluded the possibility that the GRB host metallicity aversion is caused by the decrease in galaxy metallicity with redshift.  

So while we are still dealing with small samples sizes, and all of the uncertainties that are associated with small-number statistics, all of the observations to date indicate that XRFs and GRBs arise from environments that have similar metal contents, albeit at different average redshifts.  Taken at face-value, this would imply that an additional difference is needed to explain why the jet may break out of the progenitor star of a GRB but not an XRF.

\section{Discussion: The progenitors of Ibc SNe}
\label{sec:discussion}

\subsection{Ib and Ic SNe}

The results in the preceding sections provide some vital clues to the physical and chemical properties of the progenitor stars of Ib, Ic, Ic-BL and GRB/XRF SNe.  First, it was seen that Ib SNe eject similar amounts of material as Ic SNe.  This result is perhaps unexpected if we consider that Ib and Ic SNe arise from stars of similar zero-age main sequence (ZAMS) masses.  As Ic SNe arise from stars that have been stripped of more material than Ib SNe (as the progenitor stars of Ib SNe have retained at least some of their helium envelope before exploding) logic dictates that Ib SNe should contain more mass in their ejecta.  However, this is obviously a very simplified picture, and it has been observed that some helium is present in some Ic SNe (e.g. SN 1994I showed a strong He I line at $10830$ $\rm \AA$, and SN 1987M had weak He I line at $5876$ $\rm \AA$; see Filippenko 1997 for a further review).  Moreover, \cite{Hachinger2012} have recently found that to recreate helium features in their synthetic spectra 
of Ibc SNe, only $0.06-0.14$ $\rm M_{\odot}$ of helium need to be present in the ejecta for the spectral lines to then be observed (in models where large asymmetries do not play a major role).  Therefore, only a small amount of helium needs to be present in the ejecta, and the result of \cite{Hachinger2012} naturally explains the result here that Ib and Ic SNe eject, on average, similar amounts of material.

\subsection{Ic-BL and GRB/XRF SNe}

The similarities in ejecta and nickel masses of Ic-BL and GRB/XRF SNe indicate that their progenitor stars share some common physical properties.  Additional clues to similar origins may also lie in their average and median explosion energies.  It has been noted in the literature, including a recent paper by Dessart et al. (2012), that Ic-BL SNe, including GRB/XRF SNe, are associated with larger ejecta masses and kinetic energies than Ib and Ic SNe.  Dessart et al. (2012) suggest this may be due to Ic-BL and GRB/XRF SNe arising from different progenitors to Ib and Ic SNe, and via different explosion mechanisms.

It was found here that Ic-BL are more energetic than Ib and Ic SNe, though not as energetic as GRB/XRF SNe.  As the latter arise from more metal-poor progenitors, the dividing line between Ic-BL and GRB/XRF SNe could be metal content.  Next, there is evidence that Ic-BL and GRB/XRF SNe have similar explosion mechanisms - i.e. both possess a central engine (e.g. SN 2009bb, Soderberg et al. 2010).  Thus, as the progenitors of Ic-BL SNe are more metal rich, the central engine will be less powerful than for GRB/XRF SNe, thus providing a natural explanation to the difference in their respective explosion energies.  The ejecta velocities in Table \ref{table:vel} supports this idea, as it was seen that while Ic-BL SNe have material moving at relativistic velocities, the velocities are not quite as high as in GRB/XRF SNe.

Interestingly, there is evidence that a common central engine may exist not only in Ic-BL and GRB/XRF SNe, but in \textit{all} ccSNe.  Spectroscopy of type IIn SN 2010jp (Smith et al. 2012) showed an unprecedented triple-peaked H$\alpha$ line profile, which was attributed by the authors as due to a bipolar jet-driven explosion.  It has been suggested by \cite{Chevalier2012} that ultra-luminous SNe (e.g. Gal-Yam 2012), many of which are type IIn, may arise from massive binary stars, where a SN is triggered by the in-spiral of a compact object into the core of a massive companion star, creating the theoretical Thorne-Zytkow object (Thorne \& Zytkow 1977).  Therefore the only difference between the energetic SNe that accompany GRBs and XRFs and type IIn SNe is whether the hydrogen envelope has been fully stripped before explosion.

\subsection{Progenitor Masses}

Here it was found that GRB/XRF eject more mass than Ib and Ic SNe, which may simply be due to GRB/XRF, and perhaps Ic-BL, SNe arising from more massive stars than Ib and Ic SNe.  \cite{Cano12} assembled from the literature all of the values for the ZAMS masses that were derived for individual Ibc SNe, including Ic-BL and GRB/XRF SNe (his Table 5.10 and 5.11).  It was seen that the average ZAMS mass for the stellar progenitors of Ib \& Ic, Ic-BL and GRB/XRF was 19, 30 and 35 $\rm M_{\odot}$ for sample sizes of 9, 4 and 4 respectively.  These few events would suggest that GRB/XRF and Ic-BL SNe arise from more massive stars than Ib and Ic SNe, and provides a natural explanation for why GRB/XRF and Ic-BL SNe eject more material than Ib and Ic SNe.  

Furthermore, the mass ranges of the stellar progenitors for Ic-BL and GRB/XRF SNe is close to that seen for the ZAMS mass range of WR stars in the Milky Way, which is $35-40$ $\rm M_{\odot}$ (Humphreys et al. 1985; Massey et al. 1995; Massey et al. 2003; Crowther et al. 2006, Crowther 2007).  Therefore the mass ranges for the progenitors of Ic-BL and GRB/XRF are close to that needed to form a single WR star, while those of Ib and Ic are not.

\subsection{Different Progenitor Channels}

Therefore, in terms of ejecta mass, nickel mass, explosion energies and progenitor metallicity, these results indicate that at least two progenitor channel likely exist in the production of all Ibc, including Ic-BL and GRB/XRF, SNe.  If most Ibc SNe arise from binary systems, which has been proposed by several authors (e.g. Smartt 2009; Smith et al. 2011; Eldridge et al. 2013), then one possibility of what makes Ic-BL and GRB/XRF SNe so unique is that they arise from more massive stars than Ib and Ic SNe.  These massive stars are likely to be rapidly rotating (either a rapidly rotating single star or via the in-spiral of a compact object into the core of a massive star during a common-envelope evolutionary phase), and therefore providing the necessary angular momentum needed to form an accretion disk around a central compact object.  Therefore the explosion mechanism in Ibc SNe may arise via a more ``conventional'' scenario of an iron core-collapse where the explosion is neutrino driven, while in Ic-BL and 
GRB/XRF-SNe, the explosion is driven via a central engine.  However, the possibility that Ic-BL and GRB/XRF SNe arise from in the in-spiral of a compact object into the core of a massive companion star cannot be ruled out, though it should be considered that \cite{Yoon2010} have shown that the angular momentum retained by the exploding star in the binary system is not enough to produce a magnetar nor a GRB, suggesting that GRB progenitors arise via a different process.

\section{Caveats}
\label{sec:caveats}

The analysis performed here provides a clear indication of the relative properties of Ibc SNe vs. Ic-BL and GRB/XRF SNe.  Moreover, the analysis is consistent and systematic, and does not suffer from comparing the bolometric properties of the various SNe that are derived from different methods (i.e. photometry and Arnett, or detailed modelling of spectra) and from different usages of the Arnett model.  While these results are very interesting, we must also consider the limitations of the model employed here.  

\subsection{LC Shape}

First, as was noted in \cite{Cano2011b}, SN 1998bw is not always the ideal template SN.  For example, for SN 2010bh, at early times light from the shock break-out changes the shape of the light curve, which affects how well SN~1998bw, for which no shock break-out was observed, can be used as a template.  It is worth noting that it is trivial to match the SN peak luminosities and therefore the calculated nickel masses.

\subsection{Explosion Geometry}

Next, the Arnett model used here also makes several assumptions, and an in-depth analysis of the caveats can be found here as well as in \cite{Cano12}.  One assumption includes the geometry of the explosion, which for simplicity is assumed to be spherical symmetric.  While this assumption is reasonable for Ib and Ic SNe, Ic-BL and GRB/XRF SNe are thought to be asymmetric to different degrees.  For example, if the ejecta in GRB SNe is collimated into a jet, then the amount of ejecta measured along this line of sight is not valid for all sight-lines, thus the Arnett model will over-predict the total ejecta mass.  

Furthermore, this model calculates an explosion energy by assuming that all of the ejecta are moving at the photospheric velocity, where in reality it is likely that different elements are moving at different velocities within the outflow.  Further still, the assumption of a homologous density structure that all travels at the photospheric velocity is an over-simplification of a more complex situation where ``shells'' of material (e.g. carbon, oxygen, etc.) travel at different speeds and have different densities.  Thus this assumption will also over-predict the explosion energy of the SN.

\subsection{Effective Opacity}

The model also assumes that the opacity in these events is roughly the same.  While it may be more appropriate to assume a constant opacity within a certain class of SNe (i.e. all Ib, or all Ic), the assumption of constant opacity across all classes is more debatable.  Consider Ib SNe, where the ejecta has a higher fraction of helium in it than those of Ic, Ic-BL and GRB SNe, which are rich in carbon and oxygen but deficient in helium.  However, it is expected that the overall opacity of He-rich Ib SNe and C/O-rich Ic SNe will be similar as the opacity is primarily dominated by the large number of Doppler-broadened iron lines, and it is expected that similar abundances of iron group elements exist in the ejecta of Ib and Ic SNe.  The electron scattering opacity does also contribute to the bulk opacity of the outflow however, and because it is somewhat harder to ionize helium than carbon and oxygen, the electron scattering opacity in He-rich events will be a bit lower.  However, \cite{Hachinger2012} have 
shown that only $0.06-0.14$ $\rm M_{\odot}$ of helium is required in the ejecta to be present in the observed spectra of Ib SNe, which accounts to only a small percentage of the total ejecta masses in Ibc SNe (which are of order several solar masses) and therefore may not affect the overall opacity that appreciably.

\subsection{Photospheric Velocities}

Next, an average ejecta velocity was approximated from a range of ejecta velocities, and this average velocity was used in all events in a particular SN sample that did not have a determined ejecta velocity.  This is not unusual, however, as other authors have also used average velocities when calculating the bolometric properties of Ibc SNe using an Arnett-like model, e.g. \cite{Drout2011} used a larger value of $10,000$ km s$^{-1}$, which although is larger than here, lies at the upper end of our distribution.  Nevertheless, for all of the subtypes large standard deviations were found for each of the distributions of velocities, therefore using an average value for the ejecta velocities over-simplifies the diversity in these events.  Additionally, in the literature, the peak ejecta velocities have been determined using different line diagnostics (mostly He and Fe lines; e.g. Matheson et al. 2001; Pian et al. 2006, etc.), and it therefore has not been possible to create a homogeneous sample of SNe where the 
ejecta velocity has been determined using the same spectroscopic line.

To check the effect of using an average ejecta velocity for each subtype I double checked the procedure using the following two events where the ejecta velocity is known.  Using an average photospheric velocity of $20,000 \pm 2500$ km s$^{-1}$ for SN 1998bw we find (in $UBVRIJH$): $M_{\rm ej} = 7.61$ $\rm M_{\odot}$ and $E_{\rm k} = 3.03 \times 10^{52}$ erg, which are well within the range of values in the literature: $M_{\rm ej}= 8 \pm 2$ $\rm M_{\odot}$ and $E_{\rm k} = 2-5\times 10^{52}$.  Next, for SN 2010bh (in $UBVRIJH$): $M_{\rm ej} = 1.92$ $\rm M_{\odot}$ and $E_{\rm k} = 0.77 \times 10^{52}$ erg, which is close the to published value of $M_{\rm ej}= 2.2$ $\rm M_{\odot}$, though the kinetic energy is underestimated compared to the published value ($E_{\rm k} = 1.4\times 10^{52}$) due to the lower ejecta velocity used here.

\subsection{SN Classification}

Quite importantly, and worth considering in the future, is that this paper uses the SN classification determined by the various authors in the literature.  To date no clear dividing line has been established, either observationally or theoretically, between the ejecta velocities of Ic and Ic-BL events, and while it might be safe to assume that an event with an ejecta velocity of say $\sim 8,000$ km s$^{-1}$ is a Ic SNe and another event with an ejecta velocity of $\sim 15,000$ km s$^{-1}$ is clearly a Ic-BL, events with ejecta velocities $\sim 10,000-12,000$ km s$^{-1}$ are less easily classified, and some confusion may exist in the literature as to the exact classification of a handful of events.

Moreover, for many of the historical events, the classification of the SN is made several days or more after its initial discovery.  As has been pointed out in the literature (e.g. Smartt 2009; Milisavljevic et al. 2013), some IIb events may have been misclassified as Ib SNe due to the lack of early time spectra where hydrogen lines would have been observed that then disappeared in later epochs.  Therefore it is not improbable that some of the historical Ib SNe in the sample may actually be IIb SNe.

%Therefore, in this analysis we are making the assumption that the chemical composition of the SNe are all approximately the same, and that they all have the same explosion geometries. This is obviously over-simplifying reality.  However, the Arnett model we have employed here has been used extensively by SN researchers when modelling Ibc SNe (e.g. \cite{Valenti08}; \cite{Drout2011}; \cite{Taubenberger06}; \cite{Pignata2011}).  %Moreover, the bolometric properties we have calculated for every single SN in the sample has been done in a consistent manner.

\section{Conclusions}
\label{sec:conclusion}

Here I have conducted a comprehensive and systematic analysis of the largest sample yet of Ibc SN LCs.   By using a template supernova, SN 1998bw, I was able to determine the bolometric properties (ejecta mass, nickel mass and explosion energies) of the entire sample, which where then used to obtain the general bolometric properties of Ib, Ic, Ic-BL and GRB/XRF SNe relative to each other.

Moreover, it was shown that it is possible to estimate, to an accuracy of 10\% or less, the bolometric properties of a given Ibc SN by using a template supernova, which here is SN 1998bw.  By modelling the bolometric LC of SN 1998bw, and knowledge of how much one needs to modify the SN's LC in both time (stretch) and luminosity, one is able to reproduce the overall shape of the bolometric LC for each individual SN.  Then, with knowledge of the ejecta velocity, either by modelling of spectra of the SN, or by simply using an average photospheric velocity for a given Ibc subtype, one can estimate the bolometric properties of the SN.  The validity of the method was proved by carrying out two detailed case-studies, and also from comparison of the bolometric properties determined here with those published in the literature.

Briefly, it was found that:

\begin{itemize}
 \item Ib and Ic SNe have similar amounts of ejecta mass, including nickel content thereof, and explosion energies.
 \item GRB/XRF SNe and Ic-BL SNe have similar amounts of ejecta and nickel masses.
 \item A trichotomy exists in the explosion energies of Ib \& Ic SNe vs. Ic-BL SNe vs. GRB/XRF SNe, with the latter the most energetic and the former the least energetic.
 \item GRB and XRF SNe are statistically indistinguishable.
 \item GRB and XRF SNe appear to arise from progenitors of similar metal content.
 \item Ic-BL and GRB/XRF SNe likely arise from similar types of progenitor stars, with the difference being metal content (Ic-BL are more metal rich).
 \item It is highly likely that Ic-BL and GRB/XRF SNe arise from more massive stellar progenitors than Ib and Ic SNe.
\end{itemize}

The large differences in ZAMS masses, the amount of ejecta masses, and the explosion energies between Ib \& Ic SNe and Ic-BL \& GRB/XRF SNe imply that it is likely at least two different progenitor channels exist for stripped-envelope, core-collapse supernova.  While observations support the notion that massive stars preferentially occur in binary systems, it is not unreasonable to suggest that the bulk of Ib and Ic SNe arise from the interaction of binary stars, where the primary that explodes has a mass lower than what is usually associated with WR stars.  

The progenitors of GRB/XRF SNe, and probably Ic-BL SNe, arise from either the collapse of a rapidly-rotating, massive WR star, or via common-envelope evolution of two massive stars in a binary system (which are more massive than in a ``typical'' Ibc progenitor system).  The difference between GRB/XRF SNe and Ic-BL SNe could be due to mass loss, which is governed by the metallicity of the progenitor.  As GRB/XRF SNe arise from more metal-poor progenitors than Ic-BL SNe, if we assume that rapid rotation is present in all of these events, the key difference is that the higher metallicity of the progenitors of Ic-BL SNe leads to more mass loss and crucially, more loss of angular momentum than those of GRB/XRF SNe.  It is because GRB SNe lose less mass due to their lower metal content, and hence retain more angular momentum before exploding, that they are able to power a GRB at the time of death, while Ic-BL SNe do not.  However, an additional argument is needed to explain the different origins of the gamma 
radiation in GRBs vs. XRFs, as it is seen that GRBs and XRFs arise in environments of similar metallicity.

\section{Acknowledgments}

I am deeply grateful to Cristiano Guidorzi for making his C code available for the construction of the synthetic LCs, as well as for the very useful discussions regarding the original manuscript.  I would also like to thank R. Hounsell, P. Jakobsson and G. Bj{\"o}rnsson for their insightful comments on the original manuscript.  I gratefully acknowledge support by a Project Grant from the Icelandic Research Fund.

\appendix
\label{sec:appendix}

\section{Photometry}

Details of the photometry, redshifts and extinction of the GRB/XRF SNe and the rest of the Ibc SNe can be found in Tables \ref{table:grb_photometry} and \ref{table:SNe_photometry} respectively.

\begin{table*}
\scriptsize
\centering
\setlength{\tabcolsep}{3.0pt}
\caption{Table of Photometry for the GRB/XRF SNe}
  \begin{tabular}{|ccccc|}
  \hline
GRB	&	$z$	&	$\rm E(B-V)_{fore}$	&	$\rm E(B-V)_{host}$	&	Refs.	\\
\hline

980425	&	0.0085	&	0.054	&	0.015	&	(1),(2)	\\
990712	&	0.434	&	0.027	&	0.045	&	(2--6)	\\
991208	&	0.706	&	0.015	&	0.229	&	(7),(8)	\\
011121	&	0.36	&	0.401	&	0.117	&	(8--10)	\\
020405	&	0.698	&	0.042	&	0.045	&	(2),(8),(11),(12)	\\
020903	&	0.251	&	0.027	&	0.000	&	(2),(13)	\\
021211	&	1.006	&	0.024	&	0.000	&	(14)	\\
030329	&	0.1685	&	0.021	&	0.118	&	(14)	\\
031203	&	0.1055	&	0.836	&	0.256	&	(2),(15--17)	\\
041006	&	0.716	&	0.021	&	0.033	&	(8),(18)	\\
050525A	&	0.606	&	0.075	&	0.097	&	(2),(19),(20)	\\
050824	&	0.83	&	0.035	&	0.000	&	(8),(21)	\\
060218	&	0.033	&	0.118	&	0.039	&	(2),(14),(22),(23)	\\
060729	&	0.54	&	0.033	&	0.054	&	(24)	\\
060904B	&	0.703	&	0.172	&	0.024	&	VLT archive	\\
080319B	&	0.931	&	0.009	&	0.015	&	(8),(25)	\\
090618	&	0.54	&	0.081	&	0.090	&	(24)	\\
091127	&	0.49	&	0.036	&	0.000	&	(26),(27)	\\
100316D	&	0.059	&	0.117	&	0.140	&	(28)	\\
120422A	&	0.283	&	0.037	&	0.000	&	(29),(30)	\\

\hline

\end{tabular}
%\medskip
\begin{flushleft}

(1) \cite{Galama98}, (2) \cite{Levesque10}, (3) \cite{Sahu00}, (4) \cite{Christensen04}, (5) \cite{Bjornsson01}, (6) \cite{Hjorth00}, (7) \cite{Zeh04}, (8) \cite{Kann06}, (9) \cite{Kupcu07}, (10) \cite{Garnavich03}, (11) \cite{Bersier03}, (12) \cite{Price03}, (13) \cite{Bersier06}, (14) \cite{Ferrero06}, (15) \cite{Malesani04}, (16) \cite{Mazzali2006b}, (17) \cite{Margutti07}, (18) \cite{Stanek05}, (19) \cite{DellaValle06}, (20) \cite{Blustin06}, (21) \cite{Sollerman2007}, (22) \cite{Sollerman06}, (23) \cite{Modjaz06}, (24) \cite{Cano2011a}, (25) \cite{Tanvir10}, (26) \cite{Cobb10}, (27) \cite{Berger2011}, (28) \cite{Cano2011b}, (29) \cite{Melandri2012}, (30) \cite{Schulze2013}
\end{flushleft}
\label{table:grb_photometry}

\end{table*}

\begin{table*}
\scriptsize
\centering
\setlength{\tabcolsep}{3.0pt}
\caption{Table of Photometry for the Ibc-SNe}
\label{table:SNe_photometry}

  \begin{tabular}{|cccccc|}
  \hline
SN	&	Type	&	$z$	&	$\rm E(B-V)_{fore}$	&	$\rm E(B-V)_{host}$	&	Refs.\\
\hline
1983N	&	Ib	&	0.001723	&	0.063	&	0.094	&	(1),(2),(3)	\\
1984I	&	Ib	&	0.0107	&	0.104	&	-	&	(1)	\\
1998dt	&	Ib	&	0.015	&	0.026	&	-	&	(4)	\\
1999di	&	Ib	&	0.01641	&	0.097	&	-	&	(4)	\\
1999dn	&	Ib	&	0.00938	&	0.052	&	0.048	&	(4),(5)	\\
1999ex	&	Ib	&	0.011401	&	0.019	&	0.280	&	(6)	\\
2001B	&	Ib	&	0.005227	&	0.123	&	0.000	&	(7)	\\
2004dk	&	Ib	&	0.005247	&	0.157	&	0.20	&	(8)	\\
2004gq	&	Ib	&	0.006468	&	0.073	&	0.22	&	(8)	\\
2004gv	&	Ib	&	0.019927	&	0.033	&	-	&	(8)	\\
2005az	&	Ib	&	0.008456	&	0.011	&	-	&	(8)	\\
2005bf	&	Ib pec	&	0.018913	&	0.044	&	0.000	&	(9)	\\
2005hg	&	Ib	&	0.021308	&	0.105	&	0.63	&	(8)	\\
2006dn	&	Ib	&	0.01683	&	0.113	&	-	&	(8)	\\
2006F	&	Ib	&	0.013816	&	0.19	&	-	&	(8)	\\
2007C	&	Ib	&	0.005604	&	0.042	&	0.73	&	(8)	\\
2007Y	&	Ib	&	0.004657	&	0.022	&	0.090	&	(10)	\\
2008D	&	Ib	&	0.007	&	0.022	&	0.600	&	(11)	\\
2009jf	&	Ib	&	0.007942	&	0.113	&	0.000	&	(12)	\\
1962L	&	Ic	&	0.00403	&	0.038	&	-	&	(1)	\\
1994I	&	Ic	&	0.00155	&	0.035	&	0.450	&	(13)	\\
2004aw	&	Ic	&	0.0175	&	0.364	&	0.000	&	(14)	\\
2004dn	&	Ic	&	0.012605	&	0.048	&	0.45	&	(8)	\\
2004fe	&	Ic	&	0.017895	&	0.025	&	0.23	&	(8)	\\
2004ff	&	Ic	&	0.022649	&	0.032	&	-	&	(8)	\\
2004ge	&	Ic	&	0.016128	&	0.087	&	-	&	(8)	\\
2005eo	&	Ic	&	0.017409	&	0.067	&	-	&	(8)	\\
2005mf	&	Ic	&	0.026762	&	0.018	&	0.48	&	(8)	\\
2006ab	&	Ic	&	0.01652	&	0.489	&	-	&	(8)	\\
2006fo	&	Ic	&	0.020698	&	0.029	&	-	&	(8)	\\
2007gr	&	Ic	&	0.001728	&	0.060	&	0.030	&	(15)	\\
2011bm	&	Ic	&	0.0221	&	0.032	&	0.032	&	(16)	\\
1997ef	&	Ic-BL	&	0.011693	&	0.041	&	0.174	&	(17--21)	\\
2002ap	&	Ic-BL	&	0.002187	&	0.091	&	0.000	&	(22)	\\
2003jd	&	Ic-BL	&	0.018826	&	0.044	&	0.092	&	(23),(21)	\\
2005kz	&	Ic-BL	&	0.027	&	0.054	&	-	&	(8)	\\
2007D	&	Ic-BL	&	0.023146	&	0.335	&	-	&	(8)	\\
2007ru	&	Ic-BL	&	0.01546	&	0.282	&	0.000	&	(24)	\\
2009bb	&	Ic-BL	&	0.009937	&	0.099	&	0.382	&	(25)	\\
2010ah	&	Ic-BL	&	0.0498	&	0.012	&	-	&	(26)	\\
2010ay	&	Ic-BL	&	0.067	&	0.017	&	0.200	&	(27)	\\

\hline

\end{tabular}
%\medskip
\begin{flushleft}
(1) \cite{Cadonau90}, (2) \cite{BlairPanagia1987}, (3) \cite{Clocchiatti96}, (4) \cite{Matheson01}, (5) \url{http://www.astrosurf.com/snaude/Aude/infc99dn1.htm}, (6) \cite{Stritzinger02}, (7) \url{http://www.astrosurf.com/snweb/2001/01B_/01B_Meas.htm}, (8) \cite{Drout2011}, (9) \cite{Anupama05}, (10) \cite{Stritzinger09}, (11) \cite{Malesani2009}, (12) \cite{Sahu2011}, (13) \cite{Richmond96}, (14) \cite{Taubenberger06}, (15) \cite{Hunter09}, (16) \cite{Valenti2012}, (17) \cite{Garnavich97a}, (18) \cite{Garnavich97b}, (19), \cite{Nakano1997} (20) \cite{Hu1997}, (21) \cite{Modjaz08}, (22) \cite{Yoshii2003}, (23) \cite{Valenti08}, (24) \cite{Sahu09}, (25) \cite{Pignata2011}, (26) \cite{Corsi2011}, (27) \cite{Sanders2012}	  
\end{flushleft}

\end{table*}

\section{Stretch and luminosity factors of the complete sample}

The stretch and luminosity factors of the GRB/XRF SNe and the rest of the Ibc SNe can be found in Tables \ref{table:grb_sk} and \ref{table:sn_ks} respectively.  The quoted errors are taken from GNUPlot, which are ``standard errors'' calculated by the non-linear least-squares fitting algorithm (see section \ref{sec:method}).

%Method: least squares fitting to minimise chi-squared.

\begin{landscape}
\begin{table*}

\scriptsize
\centering
\setlength{\tabcolsep}{3.0pt}
\caption{Stretch ($s$) and luminosity ($k$) factors of the GRB/XRF SNe relative to SN 1998bw}
  \begin{tabular}{|ccccccccccc|c|c|}
  \hline

GRB/XRF	&	$z$	&	$s_{B}$			&	$k_{B}$			&	$s_{V}$			&	$k_{V}$			&	$s_{R}$			&	$k_{R}$			&	$s_{I}$			&	$k_{I}$			&	$s_{\rm ave}$			&	$k_{\rm ave}$			&	Ref.	\\
  \hline
980425	&	0.0085	&		1.00				&		1.00				&		1.00				&		1.00				&		1.00				&		1.00				&		1.00				&		1.00				&		1.00				&		1.00				&	-	\\
990712	&	0.434	&		-				&		-				&	$	0.95	\pm	0.03	$	&	$	0.40	\pm	0.01	$	&	$	0.93	\pm	0.12	$	&	$	0.33	\pm	0.05	$	&		-				&		-				&	$	0.94	\pm	0.12	$	&	$	0.36	\pm	0.05	$	&	here	\\
991208	&	0.706	&		-				&		-				&		-				&		-				&	$	1.10	\pm	0.20	$	&	$	2.11	\pm	0.58	$	&		-				&		-				&	$	1.10	\pm	0.20	$	&	$	2.11	\pm	0.58	$	&	here	\\
011121	&	0.36	&		-				&		-				&		-				&		-				&	$	0.81	\pm	0.02	$	&	$	1.00	\pm	0.02	$	&		-				&		-				&	$	0.81	\pm	0.02	$	&	$	1.00	\pm	0.02	$	&	here	\\
020405	&	0.698	&		-				&		-				&		-				&		-				&	$	0.62	\pm	0.03	$	&	$	0.82	\pm	0.14	$	&		-				&		-				&	$	0.62	\pm	0.03	$	&	$	0.82	\pm	0.14	$	&	(1)	\\
020903	&	0.251	&		-				&		-				&		-				&		-				&	$	0.98	\pm	0.02	$	&	$	0.61	\pm	0.19	$	&		-				&		-				&	$	0.98	\pm	0.02	$	&	$	0.61	\pm	0.19	$	&	here	\\
021211	&	1.006	&		-				&		-				&		-				&		-				&	$	0.98	\pm	0.26	$	&	$	0.40	\pm	0.19	$	&		-				&		-				&	$	0.98	\pm	0.26	$	&	$	0.40	\pm	0.19	$	&	(2)	\\
030329	&	0.1685	&		-				&		-				&		-				&		-				&	$	0.85	\pm	0.10	$	&	$	1.50	\pm	0.19	$	&		-				&		-				&	$	0.85	\pm	0.10	$	&	$	1.50	\pm	0.19	$	&	(2)	\\
031203	&	0.1055	&		-				&		-				&		-				&		-				&	$	1.08	\pm	0.02	$	&	$	1.28	\pm	0.03	$	&	$	1.07	\pm	0.04	$	&	$	1.29	\pm	0.04	$	&	$	1.08	\pm	0.02	$	&	$	1.29	\pm	0.04	$	&	(1)	\\
041006	&	0.716	&		-				&		-				&		-				&		-				&	$	1.47	\pm	0.04	$	&	$	1.16	\pm	0.06	$	&		-				&		-				&	$	1.47	\pm	0.04	$	&	$	1.16	\pm	0.06	$	&	(1)	\\
050525A	&	0.606	&		-				&		-				&		-				&		-				&	$	0.83	\pm	0.03	$	&	$	0.69	\pm	0.03	$	&		-				&		-				&	$	0.83	\pm	0.03	$	&	$	0.69	\pm	0.03	$	&	(1)	\\
050824	&	0.83	&		-				&		-				&		-				&		-				&	$	0.52	\pm	0.14	$	&	$	1.05	\pm	0.42	$	&		-				&		-				&	$	0.52	\pm	0.14	$	&	$	1.05	\pm	0.42	$	&	(3),(4)	\\
060218$^{*}$ 	&	0.033	&	$	0.60	\pm	0.01	$	&	$	0.67	\pm	0.02	$	&	$	0.65	\pm	0.01	$	&	$	0.72	\pm	0.01	$	&	$	0.68	\pm	0.01	$	&	$	0.72	\pm	0.01	$	&	$	0.68	\pm	0.01	$	&	$	0.76	\pm	0.01	$	&	$	0.65	\pm	0.05	$	&	$	0.72	\pm	0.04	$	&	(1)	\\
060729	&	0.54	&		-				&		-				&		-				&		-				&	$	0.89	\pm	0.04	$	&	$	0.84	\pm	0.05	$	&	$	0.94	\pm	0.02	$	&	$	1.03	\pm	0.02	$	&	$	0.92	\pm	0.04	$	&	$	0.94	\pm	0.10	$	&	(1)	\\
060904B	&	0.703	&		-				&		-				&		-				&		-				&	$	0.65	\pm	0.01	$	&	$	0.48	\pm	0.02	$	&		-				&		-				&	$	0.65	\pm	0.01	$	&	$	0.42	\pm	0.02	$	&	here	\\
080319B	&	0.931	&		-				&		-				&		-				&		-				&		-				&		-				&	$	0.89	\pm	0.10	$	&	$	2.30	\pm	0.90	$	&	$	0.89	\pm	0.10	$	&	$	2.30	\pm	0.90	$	&	(1)	\\
090618	&	0.54	&		-				&		-				&		-				&		-				&	$	1.09	\pm	0.04	$	&	$	0.70	\pm	0.05	$	&	$	1.06	\pm	0.03	$	&	$	0.94	\pm	0.03	$	&	$	1.08	\pm	0.04	$	&	$	0.82	\pm	0.12	$	&	(1)	\\
091127	&	0.49	&		-				&		-				&		-				&		-				&		-				&		-				&	$	0.88	\pm	0.01	$	&	$	0.89	\pm	0.01	$	&	$	0.88	\pm	0.01	$	&	$	0.89	\pm	0.01	$	&	(1)	\\
100316D	&	0.059	&	$	0.58	\pm	0.01	$	&	$	0.41	\pm	0.01	$	&	$	0.58	\pm	0.01	$	&	$	0.43	\pm	0.01	$	&	$	0.60	\pm	0.02	$	&	$	0.40	\pm	0.01	$	&	$	0.60	\pm	0.01	$	&	$	0.48	\pm	0.01	$	&	$	0.59	\pm	0.02	$	&	$	0.43	\pm	0.05	$	&	(1)	\\
120422A	&	0.283	&		-		-		&		-				&	$	0.91	\pm	0.05	$	&	$	1.53	\pm	0.13	$	&	$	0.89	\pm	0.03	$	&	$	1.48	\pm	0.07	$	&	$	0.93	\pm	0.03	$	&	$	1.51	\pm	0.05	$	&	$	0.91	\pm	0.03	$	&	$	1.51	\pm	0.13	$	&	here	\\

\hline
\end{tabular}
%\medskip
\begin{flushleft}
\normalsize
$^{*}$  For XRF 060218/SN 2006aj, we have also determined the stretch factors in $U$: $s_{U} = 0.54 \pm 0.05$, $k_{U} = 0.73 \pm 0.03$ \\
(1) \cite{Cano2011b}, (2) \cite{Ferrero06}, (3) \cite{Sollerman2007}, (4) \cite{Kann2010}  
\end{flushleft}
\label{table:grb_sk}

\end{table*}
\end{landscape}

\begin{landscape}
\begin{table*}
\scriptsize
\centering
\setlength{\tabcolsep}{2.0pt}
\caption{Stretch ($s$) and luminosity ($k$) factors of the Ibc SNe relative to SN 1998bw}
  \begin{tabular}{|ccccccccccc|c|c|}
  \hline
SN	&	Type	&	$z$	&		$s_{B}$				&		$k_{B}$				&		$s_{V}$				&		$k_{V}$				&		$s_{R}$				&		$k_{R}$				&		$s_{I}$				&		$k_{I}$				&		$s_{\rm ave}$				&		$k_{\rm ave}$				\\
\hline

1983N	&	Ib	&	0.001723	&	$	1.08	\pm	0.13	$	&	$	0.51	\pm	0.09	$	&	$	1.13	\pm	0.03	$	&	$	0.55	\pm	0.01	$	&	$	-			$	&	$	-			$	&	$	-			$	&	$	-			$	&	$	1.11	\pm	0.13	$	&	$	0.53	\pm	0.09	$	\\
1984I	&	Ib	&	0.0107	&	$	1.07	\pm	0.09	$	&	$	0.33	\pm	0.03^{*}	$	&	$	-			$	&	$	-			$	&	$	-			$	&	$	-			$	&	$	-			$	&	$	-			$	&	$	1.07	\pm	0.09	$	&	$	0.33	\pm	0.03^{*}	$	\\
1998dt	&	Ib	&	0.015	&	$	-			$	&	$	-			$	&	$	-			$	&	$	-			$	&	$	1.00	\pm	0.08	$	&	$	0.40	\pm	0.03^{*\dag}	$	&	$	-			$	&	$	-			$	&	$	1.00	\pm	0.08	$	&	$	0.40	\pm	0.03^{*}	$	\\
1999di	&	Ib	&	0.01641	&	$	-			$	&	$	-			$	&	$	-			$	&	$	-			$	&	$	1.26	\pm	0.08	$	&	$	0.52	\pm	0.03^{*\dag}	$	&	$	-			$	&	$	-			$	&	$	1.26	\pm	0.08	$	&	$	0.52	\pm	0.03^{*}	$	\\
1999dn	&	Ib	&	0.00938	&	$	1.22	\pm	0.06	$	&	$	0.13	\pm	0.01	$	&	$	1.37	\pm	0.09	$	&	$	0.18	\pm	0.01	$	&	$	1.20	\pm	0.09	$	&	$	0.21	\pm	0.01	$	&	$	1.07	\pm	0.04	$	&	$	0.21	\pm	0.01	$	&	$	1.21	\pm	0.15	$	&	$	0.18	\pm	0.05	$	\\
1999ex	&	Ib	&	0.011401	&	$	0.93	\pm	0.01	$	&	$	0.26	\pm	0.01	$	&	$	1.03	\pm	0.02	$	&	$	0.29	\pm	0.01	$	&	$	1.15	\pm	0.02	$	&	$	0.33	\pm	0.01	$	&	$	0.82	\pm	0.01	$	&	$	0.31	\pm	0.01	$	&	$	0.98	\pm	0.17	$	&	$	0.30	\pm	0.04	$	\\
2001B	&	Ib	&	0.005227	&	$	-			$	&	$	-			$	&	$	-			$	&	$	-			$	&	$	0.99	\pm	0.06	$	&	$	0.25	\pm	0.01^{\dag}	$	&	$	-			$	&	$	-			$	&	$	0.99	\pm	0.06	$	&	$	0.25	\pm	0.01	$	\\
2004dk	&	Ib	&	0.005247	&	$	-			$	&	$	-			$	&	$	1.34	\pm	0.03	$	&	$	0.39	\pm	0.01	$	&	$	1.44	\pm	0.01	$	&	$	0.40	\pm	0.00	$	&	$	-			$	&	$	-			$	&	$	1.39	\pm	0.05	$	&	$	0.40	\pm	0.01	$	\\
2004gq	&	Ib	&	0.006468	&	$	-			$	&	$	-			$	&	$	1.03	\pm	0.02	$	&	$	0.32	\pm	0.01	$	&	$	1.02	\pm	0.01	$	&	$	0.34	\pm	0.00	$	&	$	-			$	&	$	-			$	&	$	1.02	\pm	0.02	$	&	$	0.33	\pm	0.01	$	\\
2004gv	&	Ib	&	0.019927	&	$	-			$	&	$	-			$	&	$	-			$	&	$	-			$	&	$	1.76	\pm	0.25	$	&	$	0.19	\pm	0.02^{*}	$	&	$	-			$	&	$	-			$	&	$	1.76	\pm	0.25	$	&	$	0.19	\pm	0.02^{*}	$	\\
2005az	&	Ib	&	0.008456	&	$	-			$	&	$	-			$	&	$	1.76	\pm	0.09	$	&	$	0.11	\pm	0.01^{*}	$	&	$	1.48	\pm	0.03	$	&	$	0.16	\pm	0.01^{*}	$	&	$	-			$	&	$	-			$	&	$	1.62	\pm	0.14	$	&	$	0.13	\pm	0.03^{*}	$	\\
2005bf	&	Ib pec	&	0.018913	&	$	1.22	\pm	0.15	$	&	$	0.46	\pm	0.01	$	&	$	1.30	\pm	0.16	$	&	$	0.46	\pm	0.01	$	&	$	0.95	\pm	0.23	$	&	$	0.45	\pm	0.01	$	&	$	0.83	\pm	0.17	$	&	$	0.40	\pm	0.01	$	&	$	1.08	\pm	0.23	$	&	$	0.44	\pm	0.04	$	\\
2005hg	&	Ib	&	0.021308	&	$	-			$	&	$	-			$	&	$	1.23	\pm	0.02	$	&	$	1.73	\pm	0.02	$	&	$	1.20	\pm	0.02	$	&	$	1.83	\pm	0.02	$	&	$	-			$	&	$	-			$	&	$	1.21	\pm	0.02	$	&	$	1.78	\pm	0.05	$	\\
2006F	&	Ib	&	0.013816	&	$	-			$	&	$	-			$	&	$	-			$	&	$	-			$	&	$	0.79	\pm	0.06	$	&	$	0.19	\pm	0.01^{*}	$	&	$	-			$	&	$	-			$	&	$	0.79	\pm	0.06	$	&	$	0.19	\pm	0.01^{*}	$	\\
2006dn	&	Ib	&	0.01683	&	$	-			$	&	$	-			$	&	$	-			$	&	$	-			$	&	$	1.36	\pm	0.11	$	&	$	0.19	\pm	0.01^{*}	$	&	$	-			$	&	$	-			$	&	$	1.36	\pm	0.11	$	&	$	0.19	\pm	0.01^{*}	$	\\
2007C	&	Ib	&	0.005604	&	$	-			$	&	$	-			$	&	$	0.84	\pm	0.03	$	&	$	0.53	\pm	0.01	$	&	$	0.81	\pm	0.02	$	&	$	0.52	\pm	0.01	$	&	$	-			$	&	$	-			$	&	$	0.82	\pm	0.03	$	&	$	0.52	\pm	0.01	$	\\
2007Y	&	Ib	&	0.004657	&	$	0.75	\pm	0.04	$	&	$	0.09	\pm	0.01	$	&	$	1.01	\pm	0.04	$	&	$	0.08	\pm	0.01	$	&	$	1.01	\pm	0.05^{\ddag}	$	&	$	0.08	\pm	0.01^{\ddag}	$	&	$	0.84	\pm	0.024^{\ddag}	$	&	$	0.07	\pm	0.01^{\ddag}	$	&	$	0.91	\pm	0.17	$	&	$	0.08	\pm	0.01	$	\\
2008D	&	Ib	&	0.007	&	$	1.09	\pm	0.05	$	&	$	0.13	\pm	0.01	$	&	$	1.27	\pm	0.05	$	&	$	0.15	\pm	0.01	$	&	$	1.19	\pm	0.06	$	&	$	0.16	\pm	0.01	$	&	$	1.01	\pm	0.03	$	&	$	0.18	\pm	0.01	$	&	$	1.14	\pm	0.13	$	&	$	0.16	\pm	0.02	$	\\
2009jf	&	Ib	&	0.007942	&	$	1.14	\pm	0.03	$	&	$	0.30	\pm	0.01	$	&	$	1.26	\pm	0.01	$	&	$	0.36	\pm	0.01	$	&	$	1.26	\pm	0.03	$	&	$	0.39	\pm	0.01	$	&	$	1.16	\pm	0.04	$	&	$	0.41	\pm	0.01	$	&	$	1.21	\pm	0.06	$	&	$	0.36	\pm	0.06	$	\\
1962L	&	Ic	&	0.00403	&	$	1.07	\pm	0.05	$	&	$	0.24	\pm	0.02^{*}	$	&	$	1.12	\pm	0.13	$	&	$	0.40	\pm	0.03^{*}	$	&	$	-			$	&	$	-			$	&	$	-			$	&	$	-			$	&	$	1.10	\pm	0.13	$	&	$	0.32	\pm	0.08^{*}	$	\\
1994I	&	Ic	&	0.00155	&	$	0.54	\pm	0.05	$	&	$	0.22	\pm	0.02	$	&	$	0.54	\pm	0.03	$	&	$	0.26	\pm	0.02	$	&	$	0.52	\pm	0.03	$	&	$	0.24	\pm	0.01	$	&	$	0.48	\pm	0.01	$	&	$	0.18	\pm	0.01	$	&	$	0.52	\pm	0.04	$	&	$	0.23	\pm	0.05	$	\\
2004aw	&	Ic	&	0.0175	&	$	1.07	\pm	0.06	$	&	$	0.38	\pm	0.01	$	&	$	1.25	\pm	0.02	$	&	$	0.46	\pm	0.01	$	&	$	1.18	\pm	0.02	$	&	$	0.51	\pm	0.01	$	&	$	1.07	\pm	0.02	$	&	$	0.52	\pm	0.01	$	&	$	1.16	\pm	0.16	$	&	$	0.47	\pm	0.09	$	\\
2004dn	&	Ic	&	0.012605	&	$	-			$	&	$	-			$	&	$	1.04	\pm	0.09	$	&	$	0.34	\pm	0.01	$	&	$	1.05	\pm	0.02	$	&	$	0.36	\pm	0.00	$	&	$	-			$	&	$	-			$	&	$	1.40	\pm	0.09	$	&	$	0.35	\pm	0.01	$	\\
2004fe	&	Ic	&	0.017895	&	$	-			$	&	$	-			$	&	$	0.88	\pm	0.03	$	&	$	0.53	\pm	0.01	$	&	$	0.85	\pm	0.03	$	&	$	0.52	\pm	0.01	$	&	$	-			$	&	$	-			$	&	$	0.86	\pm	0.03	$	&	$	0.53	\pm	0.01	$	\\
2004ff	&	Ic	&	0.022649	&	$	-			$	&	$	-			$	&	$	1.09	\pm	0.09	$	&	$	0.20	\pm	0.01^{*}	$	&	$	1.02	\pm	0.08	$	&	$	0.29	\pm	0.01^{*}	$	&	$	-			$	&	$	-			$	&	$	1.05	\pm	0.09	$	&	$	0.25	\pm	0.04^{*}	$	\\
2004ge	&	Ic	&	0.016128	&	$	-			$	&	$	-			$	&	$	1.10	\pm	0.21	$	&	$	0.10	\pm	0.01^{*}	$	&	$	0.96	\pm	0.11	$	&	$	0.18	\pm	0.01^{*}	$	&	$	-			$	&	$	-			$	&	$	1.03	\pm	0.21	$	&	$	0.14	\pm	0.04^{*}	$	\\
2005eo	&	Ic	&	0.017409	&	$	-			$	&	$	-			$	&	$	-			$	&	$	-			$	&	$	1.24	\pm	0.13	$	&	$	0.17	\pm	0.01^{*}	$	&	$	-			$	&	$	-			$	&	$	1.24	\pm	0.13	$	&	$	0.17	\pm	0.01^{*}	$	\\
2005mf	&	Ic	&	0.026762	&	$	-			$	&	$	-			$	&	$	1.11	\pm	0.03	$	&	$	0.87	\pm	0.02	$	&	$	0.87	\pm	0.02	$	&	$	0.84	\pm	0.01	$	&	$	-			$	&	$	-			$	&	$	0.99	\pm	0.12	$	&	$	0.86	\pm	0.02	$	\\
2006ab	&	Ic	&	0.01652	&	$	-			$	&	$	-			$	&	$	-			$	&	$	-			$	&	$	1.00	\pm	0.14	$	&	$	0.33	\pm	0.04^{*}	$	&	$	-			$	&	$	-			$	&	$	1.00	\pm	0.14	$	&	$	0.33	\pm	0.04^{*}	$	\\
2006fo	&	Ic	&	0.020698	&	$	-			$	&	$	-			$	&	$	-			$	&	$	-			$	&	$	1.17	\pm	0.28	$	&	$	0.43	\pm	0.03^{*}	$	&	$	-			$	&	$	-			$	&	$	1.17	\pm	0.28	$	&	$	0.43	\pm	0.03^{*}	$	\\
2007gr	&	Ic	&	0.001728	&	$	0.88	\pm	0.02	$	&	$	0.09	\pm	0.01	$	&	$	0.87	\pm	0.03	$	&	$	0.11	\pm	0.01	$	&	$	0.89	\pm	0.05	$	&	$	0.11	\pm	0.01	$	&	$	0.82	\pm	0.01	$	&	$	0.11	\pm	0.01	$	&	$	0.86	\pm	0.04	$	&	$	0.11	\pm	0.02	$	\\
2011bm	&	Ic	&	0.0221	&	$	2.05	\pm	0.05	$	&	$	0.49	\pm	0.01	$	&	$	2.47	\pm	0.04	$	&	$	0.60	\pm	0.00	$	&	$	2.46	\pm	0.06	$	&	$	0.77	\pm	0.01	$	&	$	1.98	\pm	0.07	$	&	$	0.82	\pm	0.01	$	&	$	2.24	\pm	0.26	$	&	$	0.67	\pm	0.15	$	\\
1997ef	&	Ic-BL	&	0.011693	&	$	-			$	&	$	-			$	&	$	1.69	\pm	0.14	$	&	$	0.31	\pm	0.01	$	&	$	-			$	&	$	-			$	&	$	-			$	&	$	-			$	&	$	1.69	\pm	0.14	$	&	$	0.31	\pm	0.01	$	\\
2002ap	&	Ic-BL	&	0.002187	&	$	0.89	\pm	0.02	$	&	$	0.19	\pm	0.01	$	&	$	0.91	\pm	0.01	$	&	$	0.29	\pm	0.00	$	&	$	0.89	\pm	0.02	$	&	$	0.30	\pm	0.01	$	&	$	0.86	\pm	0.03	$	&	$	0.25	\pm	0.01	$	&	$	0.89	\pm	0.03	$	&	$	0.26	\pm	0.07	$	\\
2003jd	&	Ic-BL	&	0.018826	&	$	0.71	\pm	0.03	$	&	$	0.90	\pm	0.03	$	&	$	0.93	\pm	0.04	$	&	$	0.89	\pm	0.02	$	&	$	0.87	\pm	0.04	$	&	$	0.87	\pm	0.02	$	&	$	0.75	\pm	0.03	$	&	$	0.95	\pm	0.02	$	&	$	0.82	\pm	0.12	$	&	$	0.00	\pm	0.05	$	\\
2005kz	&	Ic-BL	&	0.027	&	$	-			$	&	$	-			$	&	$	-			$	&	$	-			$	&	$	1.05	\pm	0.13	$	&	$	0.29	\pm	0.01^{*}	$	&	$	-			$	&	$	-			$	&	$	1.05	\pm	0.13	$	&	$	0.29	\pm	0.01^{*}	$	\\
2007D	&	Ic-BL	&	0.023146	&	$	-			$	&	$	-			$	&	$	-			$	&	$	-			$	&	$	0.78	\pm	0.09	$	&	$	0.67	\pm	0.03^{*}	$	&	$	-			$	&	$	-			$	&	$	0.78	\pm	0.09	$	&	$	0.67	\pm	0.03^{*}	$	\\
2007ru	&	Ic-BL	&	0.01546	&	$	0.64	\pm	0.02	$	&	$	0.96	\pm	0.05	$	&	$	0.86	\pm	0.03	$	&	$	1.04	\pm	0.01	$	&	$	0.79	\pm	0.06	$	&	$	1.06	\pm	0.02	$	&	$	0.68	\pm	0.02	$	&	$	1.20	\pm	0.01	$	&	$	0.74	\pm	0.11	$	&	$	1.06	\pm	0.14	$	\\
2009bb	&	Ic-BL	&	0.009937	&	$	0.68	\pm	0.02	$	&	$	0.52	\pm	0.01	$	&	$	0.79	\pm	0.01	$	&	$	0.65	\pm	0.01	$	&	$	0.77	\pm	0.05	$	&	$	0.62	\pm	0.01	$	&	$	0.67	\pm	0.03	$	&	$	0.62	\pm	0.01	$	&	$	0.73	\pm	0.07	$	&	$	0.60	\pm	0.05	$	\\
2010ah	&	Ic-BL	&	0.0498	&	$	-			$	&	$	-			$	&	$	-			$	&	$	-			$	&	$	0.94	\pm	0.04	$	&	$	0.49	\pm	0.02^{*}	$	&	$	-			$	&	$	-			$	&	$	0.94	\pm	0.04	$	&	$	0.49	\pm	0.02^{*}	$	\\
2010ay	&	Ic-BL	&	0.067	&	$	-			$	&	$	-			$	&	$	-			$	&	$	-			$	&	$	0.99	\pm	0.13	$	&	$	2.42	\pm	0.41	$	&	$	-			$	&	$	-			$	&	$	0.99	\pm	0.13	$	&	$	2.42	\pm	0.41	$	\\

\hline
\end{tabular}
\label{table:sn_ks}
\begin{flushleft}
\normalsize
$^{*}$ The rest-frame extinction is not known for these events, so the luminosity factors computed here can be regarded as lower limits. \\
$^{\dag}$ Observations were made in the clear filter, which closely approximates to $R$. \\
$^{\ddag}$ For SN 2007Y, the stretch factors in R- and I-band are in SDSS filters $r$ and $i$ respectively.\\
\end{flushleft}

\end{table*}
\end{landscape}

\section{Bolometric properties of the complete sample}

The bolometric properties of the GRB/XRF SNe and the rest of the Ibc SNe can be found in Tables \ref{table:grb_bol} and \ref{table:sn_bol} respectively.  The quoted errors have been computed by considering the maximum and minimum stretch and luminosity factors, as well as the distribution of velocities around the average value (for those events where an average ejecta velocity is used).

\begin{landscape}

\begin{table*}
%\scriptsize
\centering
\setlength{\tabcolsep}{2pt}
\caption{Bolometric properties of the GRB/XRF SNe in $UBVRIJH$}
  \begin{tabular}{|ccccccc|cccc|}
  \hline
GRB	&	$z$	&	$v_{\rm ph}$ (km s$^{-1}$)	&		$M_{\rm Ni}$ ($\rm M_{\odot}$)				&		$M_{\rm ej}$ ($\rm M_{\odot}$)				&		$E_{\rm k}$ $(10^{52}$ erg)				&	$E_{\rm k}/M_{\rm ej}$ 	&	$M_{\rm Ni}$ ($\rm M_{\odot}$)$^{\zeta}$	&	$M_{\rm ej}$ ($\rm M_{\odot}$)$^{\zeta}$	&	$E_{\rm k}$ $(10^{52}$ erg)$^{\zeta}$	&	Ref.	\\

\hline
980425	&	0.0085	&	18,000 	&	$	0.42	\pm	0.02	$	&	$	6.80	\pm	0.57	$	&	$	2.19	\pm	0.17	$	&	0.322	&	$0.4-0.7$	&	$8 \pm 2$	&	$2-5$	&	(1),(2),(3)	\\
990712	&	0.434	&	\textit{20,000}	&	$	0.14	\pm	0.04	$	&	$	6.55	^{+3.52}	_{-2.92}	$	&	$	2.61	^{+2.46}	_{-1.50}	$	&	0.398	&	-	&	-	&	-	&	-	\\
991208	&	0.706	&	\textit{20,000}	&	$	0.96	\pm	0.48	$	&	$	9.72	^{+6.81}	_{-5.58}	$	&	$	3.87	^{+4.46}	_{-2.60}	$	&	0.398	&	-	&	-	&	-	&	-	\\
011121	&	0.36	&	\textit{20,000}	&	$	0.35	\pm	0.01	$	&	$	4.44	^{+0.82}	_{-0.77}	$	&	$	1.77	^{+0.88}	_{-0.64}	$	&	0.398	&	-	&	-	&	-	&	-	\\
020405	&	0.698	&	\textit{20,000}	&	$	0.23	\pm	0.02	$	&	$	2.24	^{+0.61}	_{-0.53}	$	&	$	0.89	^{+0.54}	_{-0.38}	$	&	0.398	&	-	&	-	&	-	&	-	\\
020903	&	0.251	&	\textit{20,000}	&	$	0.25	\pm	0.13	$	&	$	7.27	^{+4.88}	_{-4.00}	$	&	$	2.89	^{+3.22}	_{-1.89}	$	&	0.398	&	-	&	-	&	-	&	-	\\
021211	&	1.006	&	\textit{20,000}	&	$	0.16	\pm	0.14	$	&	$	7.16	^{+7.44}	_{-5.99}	$	&	$	2.85	^{+4.50}	_{-1.30}	$	&	0.398	&	-	&	-	&	-	&	-	\\
030329	&	0.1685	&	20,000 	&	$	0.54	\pm	0.13	$	&	$	5.06	\pm	1.65	$	&	$	2.01	\pm	0.65	$	&	0.398	&	0.4	&	$7 \pm 3$	&	$3.5 \pm 1.5$	&	(4)	\\
031203	&	0.1055	&	18,000 	&	$	0.57	\pm	0.04	$	&	$	8.22	\pm	0.76	$	&	$	2.65	\pm	0.25	$	&	0.322	&	0.55	&	13	&	6	&	(5)	\\
041006	&	0.716	&	\textit{20,000}	&	$	0.69	\pm	0.07	$	&	$	19.20	^{+3.87}	_{-3.55}	$	&	$	7.64	^{+3.98}	_{-2.87}	$	&	0.398	&	-	&	-	&	-	&	-	\\
050525A	&	0.606	&	\textit{20,000}	&	$	0.24	\pm	0.02	$	&	$	4.75	^{+1.08}	_{-1.00}	$	&	$	1.89	^{+1.07}	_{-0.75}	$	&	0.398	&	-	&	-	&	-	&	-	\\
050824	&	0.83	&	\textit{20,000}	&	$	0.26	\pm	0.17	$	&	$	1.44	^{+1.56}	_{-0.64}	$	&	$	0.57	^{+0.93}	_{-0.37}	$	&	0.398	&	-	&	-	&	-	&	-	\\
060218	&	0.033	&	20,000 	&	$	0.21	\pm	0.03	$	&	$	2.58	\pm	0.55	$	&	$	1.02	\pm	0.23	$	&	0.395	&	0.2	&	2	&	0.2	&	(6)	\\
060729	&	0.54	&	\textit{20,000}	&	$	0.36	\pm	0.05	$	&	$	6.13	^{+1.56}	_{-1.39}	$	&	$	2.44	^{+1.43}	_{-0.99}	$	&	0.398	&	-	&	-	&	-	&	-	\\
060904B	&	0.703	&	\textit{20,000}	&	$	0.12	\pm	0.01	$	&	$	2.50	^{+0.50} 	_{-0.47}	$	&	$	0.99	^{+0.51}	_{-0.37}	$	&	0.398	&	0.1	&	-	&	-	&	(7)	\\
080319B	&	0.931	&	\textit{20,000}	&	$	0.86	\pm	0.45	$	&	$	5.72	^{+2.59}	_{-2.18}	$	&	$	2.27	^{+1.91}	_{-1.19}	$	&	0.398	&	-	&	-	&	-	&	-	\\
090618	&	0.54	&	\textit{20,000}	&	$	0.37	\pm	0.03	$	&	$	9.17	^{+2.06}	_{-1.85}	$	&	$	3.65	^{+2.00}	_{-1.42}	$	&	0.398	&	-	&	-	&	-	&	-	\\
091127	&	0.49	&	17,000 	&	$	0.33	\pm	0.01	$	&	$	4.69	\pm	0.13	$	&	$	1.35	\pm	0.04	$	&	0.287	&	0.35	&	1.4	&	0.2	&	(8)	\\
100316D	&	0.059	&	25,000 	&	$	0.12	\pm	0.02	$	&	$	2.47	\pm	0.23	$	&	$	1.54	\pm	0.14	$	&	0.623	&	0.1	&	2.2	&	1.4	&	(9)	\\
120422A	&	0.283	&	20,500 	&	$	0.57	\pm	0.07	$	&	$	6.10	\pm	0.49	$	&	$	2.55	\pm	0.21	$	&	0.418	&	0.58	&	5.87	&	4.1	&	(10),(11)	\\

\hline
\end{tabular}
\label{table:grb_bol}
\begin{flushleft}
$^{\zeta}$ \textbf{Literature values.}\\
$^{*}$  The photospheric velocity in italics is an \textit{average} velocity for GRB SNe (see Table \ref{table:vel}), and the value $v_{\rm ph} = 20,000 \pm 2,500$ km s$^{-1}$ is used when calculating the the ejecta masses and kinetic energies, as well as their corresponding errors.  Velocities in plain font have been derived from their respective spectra. \\
(1) \cite{Patat01}, (2) \cite{Iwamoto98}, (3) \cite{Nakamura01}, (4) \cite{Deng05}, (5) \cite{Mazzali2006b}, (6) \cite{Mazzali2006a}, (7) \cite{Margutti2008}, (8) \cite{Berger2011}, (9) \cite{Cano2011b}, (10) \cite{Melandri2012}, (11) \cite{Schulze2013}
\end{flushleft}
\end{table*}
\end{landscape}

\begin{landscape}
\begin{table*}
\scriptsize
\centering
\setlength{\tabcolsep}{2pt}
\caption{Bolometric properties of the Ibc SNe in $UBVRIJH$}
  \begin{tabular}{|cccccccc|cccc|}
  \hline
SN	&	Type	&	$z$	&	$v_{\rm ph}$ (km s$^{-1}$)$^{*}$	&		$M_{\rm Ni}$ ($\rm M_{\odot}$)				&		$M_{\rm ej}$ ($\rm M_{\odot}$)				&		$E_{\rm k}$ $(10^{52}$ erg)				&	$E_{\rm k}/M_{\rm ej}$ 	&	$M_{\rm Ni}$ ($\rm M_{\odot}$)$^{\zeta}$	&	$M_{\rm ej}$ ($\rm M_{\odot}$)$^{\zeta}$	&	$E_{\rm k}$ $(10^{52}$ erg)$^{\zeta}$	&	Ref.	\\
\hline
1983N	&	Ib	&	0.001723	&	\textit{8,000}	&	$	0.24	\pm	0.07	$	&	$	3.89	^{+2.52}	_{-1.90}	$	&	$	0.25	^{+0.36}	_{-0.21}	$	&	0.064	&	0.16	&	4.5	&	0.27	&	(1)	\\
1984I	&	Ib	&	0.0107	&	\textit{8,000}	&	$	0.15	\pm	0.03^{\dag}	$	&	$	3.62	^{+1.81}	_{-1.46}	$	&	$	0.23	^{+0.31}	_{-0.15}	$	&	0.064	&	-	&	-	&	-	&	-	\\
1998dt	&	Ib	&	0.015	&	7,200	&	$	0.17	\pm	0.03^{\dag}	$	&	$	2.70	\pm	0.62	$	&	$	0.14	\pm	0.032	$	&	0.051	&	-	&	-	&	-	&	-	\\
1999di	&	Ib	&	0.01641	&	6,500	&	$	0.27	\pm	0.03^{\dag}	$	&	$	4.36	\pm	0.74	$	&	$	0.18	\pm	0.03	$	&	0.042	&	-	&	-	&	-	&	-	\\
1999dn	&	Ib	&	0.00938	&	8,500	&	$	0.09	\pm	0.04	$	&	$	5.13	\pm	1.82	$	&	$	0.37	\pm	0.13	$	&	0.072	&	0.11	&	$4-6$	&	0.5	&	(2)	\\
1999ex	&	Ib	&	0.011401	&	\textit{8,000}	&	$	0.12	\pm	0.04	$	&	$	2.91	^{+2.50}	_{-1.80}	$	&	$	0.19	^{+0.35}	_{-0.15}	$	&	0.064	&	0.16	&	4.5	&	0.27	&	(3)	\\
2001B	&	Ib	&	0.005227	&	\textit{8,000}	&	$	0.10	\pm	0.01	$	&	$	2.86	^{+1.33}	_{-1.08}	$	&	$	0.18	^{+0.23}	_{-0.11}	$	&	0.064	&	-	&	-	&	-	&	-	\\
2004dk	&	Ib	&	0.005247	&	\textit{8,000}	&	$	0.23	\pm	0.01	$	&	$	6.82	^{+2.41}	_{-2.14}	$	&	$	0.43	^{+0.48}	_{-0.26}	$	&	0.064	&	0.23	&	3.61	&	0.23	&	(4)	\\
2004gq	&	Ib	&	0.006468	&	\textit{8,000}	&	$	0.14	\pm	0.01	$	&	$	3.19	^{+1.03}	_{-0.94}	$	&	$	0.20	^{+0.21}	_{-0.12}	$	&	0.064	&	0.13	&	2.11	&	0.13	&	(4)	\\
2004gv	&	Ib	&	0.019927	&	\textit{8,000}	&	$	0.14	\pm	0.04^{\dag}	$	&	$	11.72	^{+8.25}	_{-6.26}	$	&	$	0.75	^{+1.23}	_{-0.54}	$	&	0.064	&	-	&	-	&	-	&	-	\\
2005az	&	Ib	&	0.008456	&	\textit{8,000}	&	$	0.09	\pm	0.02^{\dag}	$	&	$	10.22	^{+4.21}	_{-3.54}	$	&	$	0.65	^{+0.78}	_{-0.41}	$	&	0.064	&	0.31	&	4.05	&	0.25	&	(4)	\\
2005bf	&	Ib pec	&	0.018913	&	6,000	&	$	0.20	\pm	0.06	$	&	$	2.71	\pm	1.63	$	&	$	0.10	\pm	0.06	$	&	0.036	&	0.31	&	$6-7$	&	$0.1-0.2$	&	(5),(6)	\\
2005hg	&	Ib	&	0.021308	&	\textit{8,000}	&	$	0.89	\pm	0.04	$	&	$	4.92	^{+1.45}	_{-1.36}	$	&	$	0.31	^{+0.32}	_{-0.18}	$	&	0.064	&	0.64	&	2.11	&	0.13	&	(4)	\\
2006dn	&	Ib	&	0.01683	&	\textit{8,000}	&	$	0.10	\pm	0.01^{\dag}	$	&	$	6.41	^{+3.37}	_{-2.66}	$	&	$	0.41	^{+0.56}	_{-0.27}	$	&	0.064	&	0.30	&	3.61	&	0.23	&	(4)	\\
2006F	&	Ib	&	0.013816	&	\textit{8,000}	&	$	0.06	\pm	0.01^{\dag}	$	&	$	1.64	^{+0.82}	_{-0.65}	$	&	$	0.10	^{+0.14}	_{-0.07}	$	&	0.063	&	0.21	&	1.51	&	0.09	&	(4)	\\
2007C	&	Ib	&	0.005604	&	\textit{8,000}	&	$	0.18	\pm	0.01	$	&	$	1.83	^{+0.71}	_{-0.61}	$	&	$	0.12	^{+0.13}	_{-0.07}	$	&	0.063	&	0.16	&	1.51	&	0.09	&	(4)	\\
2007Y	&	Ib	&	0.004657	&	7,000	&	$	0.03	\pm	0.01	$	&	$	2.09	\pm	0.95	$	&	$	0.10	\pm	0.05	$	&	0.048	&	0.06	&	0.42	&	0.01	&	(7)	\\
2008D	&	Ib	&	0.007	&	10,000	&	$	0.08	\pm	0.02	$	&	$	5.33	\pm	1.34	$	&	$	0.53	\pm	0.13	$	&	0.099	&	0.09	&	7	&	0.6	&	(8)	\\
2009jf	&	Ib	&	0.007942	&	12,000	&	$	0.18	\pm	0.04	$	&	$	7.34	\pm	1.00	$	&	$	1.05	\pm	0.15	$	&	0.143	&	$0.17 \pm 0.03$	&	$4-9$	&	$0.3-0.8$	&	(9)	\\
1962L	&	Ic	&	0.00403	&	\textit{8,000}	&	$	0.15	\pm	0.04^{\dag}	$	&	$	3.73	^{+1.33}	_{-1.17}	$	&	$	0.24	^{+0.26}	_{-0.09}	$	&	0.064	&	-	&	-	&	-	&	-	\\
1994I	&	Ic	&	0.00155	&	10,000	&	$	0.06	\pm	0.01	$	&	$	0.72	\pm	0.04	$	&	$	0.07	\pm	0.01	$	&	0.100	&	0.07	&	0.9	&	0.1	&	(10)	\\
2004aw	&	Ic	&	0.0175	&	11,800	&	$	0.22	\pm	0.08	$	&	$	6.49	\pm	2.32	$	&	$	0.90	\pm	0.32	$	&	0.139	&	$0.2-0.3$	&	$3-8$	&	0.8	&	(11)	\\
2004dn	&	Ic	&	0.012605	&	\textit{8,000}	&	$	0.15	\pm	0.02	$	&	$	3.40	^{+1.79}	_{-1.42}	$	&	$	0.22	^{+0.30}	_{-0.14}	$	&	0.064	&	0.16	&	3.61	&	0.23	&	(4)	\\
2004fe	&	Ic	&	0.017895	&	\textit{8,000}	&	$	0.19	\pm	0.01	$	&	$	2.07	^{+0.80}	_{-0.67}	$	&	$	0.13	^{+0.15}	_{-0.08}	$	&	0.064	&	0.19	&	1.25	&	0.08	&	(4)	\\
2004ff	&	Ic	&	0.022649	&	\textit{8,000}	&	$	0.11	\pm	0.03^{\dag}	$	&	$	3.50	^{+1.82}	_{-1.43}	$	&	$	0.22	^{+0.30}	_{-0.15}	$	&	0.064	&	0.18	&	1.25	&	0.08	&	(4)	\\
2004ge	&	Ic	&	0.016128	&	\textit{8,000}	&	$	0.06	\pm	0.03^{\dag}	$	&	$	3.21	^{+3.13}	_{-2.19}	$	&	$	0.20	^{+0.42}	_{-0.17}	$	&	0.064	&	0.59	&	2.45	&	0.15	&	(4)	\\
2005eo	&	Ic	&	0.017409	&	\textit{8,000}	&	$	0.09	\pm	0.01^{\dag}	$	&	$	5.13	^{+2.88}	_{-2.24}	$	&	$	0.33	^{+0.47}	_{-0.22}	$	&	0.064	&	0.33	&	1.51	&	0.09	&	(4)	\\
2005mf	&	Ic	&	0.026762	&	\textit{8,000}	&	$	0.35	\pm	0.05	$	&	$	2.98	^{+2.01}	_{-1.51}	$	&	$	0.19	^{+0.31}	_{-0.13}	$	&	0.063	&	-	&	-	&	-	&	-	\\
2006ab	&	Ic	&	0.01652	&	\textit{8,000}	&	$	0.14	\pm	0.04^{\dag}	$	&	$	3.02	^{+2.31}	_{-1.68}	$	&	$	0.19	^{+0.34}	_{-0.14}	$	&	0.064	&	0.23	&	2.45	&	0.15	&	(4)	\\
2006fo	&	Ic	&	0.020698	&	\textit{8,000}	&	$	0.21	\pm	0.07^{\dag}	$	&	$	4.40	^{+4.97}	_{-3.42}	$	&	$	0.28	^{+0.65}	_{-0.24}	$	&	0.064	&	-	&	-	&	-	&	-	\\
2007gr	&	Ic	&	0.001728	&	6,700	&	$	0.04	\pm	0.01	$	&	$	1.70	\pm	0.23	$	&	$	0.08	\pm	0.01	$	&	0.045	&	$0.076 \pm 0.010$	&	$2-3.5$	&	$0.1-0.4$	&	(12)	\\
2011bm	&	Ic	&	0.0221	&	8,000	&	$	0.58	\pm	0.19	$	&	$	18.75	\pm	3.74	$	&	$	1.19	\pm	0.24	$	&	0.063	&	$0.6-0.7$	&	$7-17$	&	$0.7-1.7$	&	(13)	\\
1997ef	&	Ic-BL	&	0.011693	&	10,000	&	$	0.21	\pm	0.01	$	&	$	13.32	\pm	2.09	$	&	$	1.32	\pm	0.20	$	&	0.099	&	$0.15 \pm 0.03$	&	10	&	0.8	&	(14)	\\
2002ap	&	Ic-BL	&	0.002187	&	14,000	&	$	0.10	\pm	0.29	$	&	$	3.90	\pm	0.28	$	&	$	0.76	\pm	0.05	$	&	0.195	&	0.07	&	$2.5-5.0$	&	$0.4-1.0$	&	(15)	\\
2003jd	&	Ic-BL	&	0.018826	&	13,500	&	$	0.31	\pm	0.04	$	&	$	3.07	\pm	1.22	$	&	$	0.56	\pm	0.22	$	&	0.181	&	$0.36 \pm 0.04$	&	$3.0 \pm 0.5$	&	0.7	&	(16)	\\
2005kz	&	Ic-BL	&	0.027	&	\textit{15,000}	&	$	0.12	\pm	0.02^{\dag}	$	&	$	6.52	^{+4.41}	_{-3.29}	$	&	$	1.46	^{+2.46}	_{-1.07}	$	&	0.224	&	0.47	&	8.1	&	2.03	&	(4)	\\
2007D	&	Ic-BL	&	0.023146	&	\textit{15,000}	&	$	0.23	\pm	0.03^{\dag}	$	&	$	3.10	^{+2.05}	_{-1.54}	$	&	$	0.69	^{+1.15}	_{-0.50}	$	&	0.224	&	1.50	&	3.6	&	0.9	&	(4)	\\
2007ru	&	Ic-BL	&	0.01546	&	20,000	&	$	0.34	\pm	0.09	$	&	$	3.59	\pm	1.52	$	&	$	1.43	\pm	0.57	$	&	0.398	&	0.33	&	$1.3 \pm 1.1$	&	$0.50-0.86$	&	(17)	\\
2009bb	&	Ic-BL	&	0.009937	&	15,000	&	$	0.19	\pm	0.03	$	&	$	2.55	\pm	0.06	$	&	$	0.57	\pm	0.13	$	&	0.224	&	$0.22 \pm 0.06$	&	$4.1 \pm 1.9$	&	$1.8 \pm 0.07$	&	(18)	\\
2010ah	&	Ic-BL	&	0.0498	&	\textit{15,000}	&	$	0.19	\pm	0.01^{\dag}	$	&	$	4.87	^{+1.88}	_{-1.64}	$	&	$	1.09	^{+1.33}	_{-0.70}	$	&	0.224	&	$0.2-0.25$	&	$6 \pm 2$	&	$1.5 \pm 0.5$	&	(19)	\\
2010ay	&	Ic-BL	&	0.067	&	21,000	&	$	1.00	\pm	0.31	$	&	$	7.84	\pm	2.71	$	&	$	3.44	\pm	1.19	$	&	0.439	&	$0.9 \pm 0.1$	&	$>4.7$	&	$>1.08$	&	(20)	\\

\hline
\end{tabular}
\label{table:sn_bol}
\begin{flushleft}
%\scriptsize
$^{\zeta}$ \textbf{Literature values.}\\
$^{*}$  The photospheric velocities in italics are \textit{average} velocities for the Ibc SNe (see Table \ref{table:vel}).  The value $v_{\rm ph} = 8,000 \pm 2,000$ km s$^{-1}$ is used for Ib and Ic SNe, and $v_{\rm ph} = 15,000 \pm 4,000$ km s$^{-1}$ for Ic-BL SNe when calculating the ejecta masses and kinetic energies, as well as their corresponding errors.  Velocities in plain font have been derived from their respective spectra. \\
$^{\dag}$  The rest-frame extinction is not known for these events, so the nickel masses computed here can be regarded as lower limits.  These values are \textit{not} considered in the statistical analyses (section \ref{sec:stats_analysis}).\\
(1) \cite{EnsmanWoos1988}, (2) \cite{Benetti2011}, (3) \cite{Stritzinger02}, (4) \cite{Drout2011}, (5) \cite{Folatelli2006}, (6) \cite{Anupama05}, (7) \cite{Stritzinger09}, (8) \cite{Mazzali08}, (9) \cite{Sahu2011}, (10) \cite{Iwamoto94}, (11) \cite{Taubenberger06}, (12) \cite{Hunter09}, (13) \cite{Valenti2012}, (14) \cite{Iwamoto2000}, (15) \cite{Mazzali2002}, (16) \cite{Valenti08}, (17) \cite{Sahu09}, (18) \cite{Pignata2011}, (19) \cite{Corsi2011}, (20) \cite{Sanders2012}
\end{flushleft}
\end{table*}
\end{landscape}

%Refs: (1), (2), (3), (4), (5), (6), (7), (8), (9), (10), (11), (12), (13), (14), (15), (16), (17), (18), (19), (20)

\section{Photospheric velocities of the Ibc SNe}

The photospheric velocities of the Ibc SNe are displayed in Table \ref{table:vel}.

\begin{landscape}
\begin{table*}
\setlength{\tabcolsep}{5pt}
 \centering
 %\begin{minipage}{220mm}
  \caption{Photospheric velocities of Ibc SNe.}
  \begin{tabular}{|cccccccc|}
  \hline
SN	&	Type	&	$v_{\rm ph}$ (km s$^{-1}$)	&	Ref.	&	SN	&	Type	&	$v_{\rm ph}$ (km s$^{-1}$)	&	Ref.	\\
\hline															
1984L	&	Ib	&	7200	&	(1)	&	2004aw	&	Ic	&	11800	&	(7)	\\
1991ar	&	Ib	&	8000	&	(1)	&	2007gr	&	Ic	&	6700	&	(8)	\\
1997dc	&	Ib	&	8200	&	(1)	&	2011bm	&	Ic	&	8000	&	(9)	\\
1998I	&	Ib	&	7700	&	(1)	&	1997dq	&	Ic-BL	&	12300	&	(1)	\\
1998dt	&	Ib	&	7200	&	(1)	&	1997ef	&	Ic-BL	&	10000	&	(10)	\\
1999di	&	Ib	&	6500	&	(1)	&	2002ap	&	Ic-BL	&	14000	&	(11)	\\
1999dn	&	Ib	&	8500	&	(1)	&	2003jd	&	Ic-BL	&	13500	&	(12)	\\
2005bf	&	Ib pec	&	6000	&	(2)	&	2007ru	&	Ic-BL	&	20000	&	(13)	\\
2007Y	&	Ib	&	7000	&	(3)	&	2009bb	&	Ic-BL	&	15000	&	(14)	\\
2008D	&	Ib	&	10000	&	(4)	&	2010ay	&	Ic-BL	&	21000	&	(15)	\\
2009jf	&	Ib	&	12000	&	(5)	&	1998bw	&	GRB/XRF	&	18000	&	(16)	\\
1988L	&	Ic	&	6400	&	(1)	&	2003dh	&	GRB/XRF	&	20000	&	(17)	\\
1990B	&	Ic	&	7400	&	(1)	&	2003lw	&	GRB/XRF	&	18000	&	(18)	\\
1990U	&	Ic	&	6700	&	(1)	&	2006aj	&	GRB/XRF	&	20000	&	(19)	\\
1990aa	&	Ic	&	8300	&	(1)	&	2009nz	&	GRB/XRF	&	17000	&	(20)	\\
1994I	&	Ic	&	10000	&	(6)	&	2010bh	&	GRB/XRF	&	25000	&	(21)	\\
1995F	&	Ic	&	9700	&	(1)	&	2012bz	&	GRB/XRF	&	20500	&	(22)	\\
1997ei	&	Ic	&	9700	&	(1)	&	-	&	-	&	-	&	-	\\
\hline															
-	&	SN type	&	$v_{\rm ph,ave}$ (km s$^{-1}$)	&	$\sigma$ (km s$^{-1}$)	&	$v_{\rm ph,median}$ (km s$^{-1}$)	&	N	&	$v_{\rm ph,model}$ (km s$^{-1}$)	&	-	\\
\hline															
-	&	Ib	&	8027	&	1700	&	7700	&	11	&	$8000 \pm 2000$	&	-	\\
-	&	Ic	&	8470	&	1776	&	8150	&	10	&	$8000 \pm 2000$	&	-	\\
-	&	Ic-BL	&	15114	&	4009	&	14000	&	7	&	$15,000 \pm 4000$	&	-	\\
-	&	GRB	&	19786	&	2644	&	20000	&	7	&	$20,000 \pm 2500$	&	-	\\
\hline
\end{tabular}
\label{table:vel}

\begin{flushleft}
%\scriptsize
(1) \cite{Matheson01}, (2) \cite{Anupama05}, (3) \cite{Stritzinger09}, (4) \cite{Mazzali08}, (5) \cite{Sahu2011}, (6) \cite{Richmond96}, (7) \cite{Taubenberger06}, (8) \cite{Hunter09}, (9) \cite{Valenti2012}, (10) \cite{Mazzali2000}, (11) \cite{Mazzali2002}, (12) \cite{Valenti08}, (13) \cite{Sahu09}, (14) \cite{Pignata2011}, (15) \cite{Sanders2012}, (16) \cite{Galama98}, (17) \cite{Matheson03}, (18) \cite{Mazzali2006b}, (19) \cite{Pian06}, (20) \cite{Berger2011}, (21) \cite{Chornock10}, (22) \cite{Schulze2013}
\end{flushleft}

%\end{minipage}
\end{table*}
\end{landscape}
%\bsp

\label{lastpage}

\end{document}